\documentclass[acmsmall, screen=true, review=false]{acmart}

\AtBeginDocument{%
  \providecommand\BibTeX{{%
    \normalfont B\kern-0.5em{\scshape i\kern-0.25em b}\kern-0.8em\TeX}}}

\usepackage{tikz}
\usetikzlibrary{mindmap,shapes,positioning,shadows}
\usepackage{forest}
\usepackage{smartdiagram}
\usesmartdiagramlibrary{additions}
\usepackage{graphicx}
\usepackage{subcaption}
\usepackage{amsmath,amsfonts,mathtools}
\usepackage{amsthm}          % For Theorems & Definitions

\newtheorem{myDef}{Definition}

\usepackage{tcolorbox}
\usepackage{bbding}          % 引入宏包
\usepackage{placeins}        % for \FloatBarrier
\usepackage{algorithm}
\usepackage[noend]{algpseudocode}

\usepackage{hyperref}
\usepackage{cleveref} 
\usepackage{booktabs}
\usepackage{siunitx}
\usepackage{etoolbox}
\usepackage{nicematrix}
\usepackage{multirow}
\usepackage{multicol}
\usepackage{makecell}
\usepackage{pifont}
\usepackage{enumitem}
\usepackage{fontawesome5}
\usepackage{array}
\usepackage{geometry}
\usepackage{graphicx}
\usepackage{ragged2e} % for \raggedright
\usepackage{diagbox}  % for diagonal box
\usepackage{mathrsfs} % For Symbols
\usepackage{stfloats}
\usepackage{url}
\usepackage{textcomp}
\usepackage{xcolor}
\usepackage{hhline}          % 双线表 
\usepackage{threeparttable} 
\usepackage{float}
\usepackage{colortbl}
\usepackage{soul}

\usepackage{tabularx}
\usepackage{threeparttable}
\usepackage{lscape}          % For landscape
\usepackage{adjustbox}       % To adjust table width

\hypersetup{colorlinks=true, citecolor=blue}
\usepackage[T1]{fontenc}
\usepackage[utf8]{inputenc}
\usepackage[english]{babel}
\usepackage{natbib}

\crefname{figure}{Fig.}{Figs.}
\Crefname{figure}{Fig.}{Figs.}

\crefname{table}{Table}{Tables}
\Crefname{table}{Table}{Tables}

\crefname{section}{Sec.}{Secs.}
\Crefname{section}{Sec.}{Secs.}
\crefname{equation}{Eq.}{Eqs.}
\Crefname{equation}{Eq.}{Eqs.}

\usepackage{cellspace}
\begin{document}

\title{Secure and Robust Watermarking for AI-generated Images: A Comprehensive Survey}
% \title{Text Watermarking in Large Language Models: A Survey}

\author{Jie Cao}
\email{jie.cao@queensu.ca}
\affiliation{
  \institution{Department of Electrical and Computer Engineering, Queen's University}
  \city{Kingston}
  \country{Canada}
  }
  
\author{Qi Li}
\email{qi.li@queensu.ca}
\affiliation{
  \institution{School of Computing, Queen's University}
  \city{Kingston}
  \country{Canada}
  }

\author{Zelin Zhang}
\email{zelin.zhang@queensu.ca}
\affiliation{
  \institution{Department of Electrical and Computer Engineering, Queen's University}
  \city{Kingston}
  \country{Canada}
  }

  \author{Jianbing Ni}
\email{jianbing.ni@queensu.ca}
  \authornote{Corresponding author.}
\affiliation{
  \institution{Department of Electrical and Computer Engineering, Queen's University}
  \city{Kingston}
  \country{Canada}  }

  \author{Rongxing Lu}
\email{rongxing.lu@queensu.ca}
\affiliation{
  \institution{School of Computing, Queen's University}
  \city{Kingston}
  \country{Canada}
  }

\renewcommand{\shortauthors}{Cao, et al.}

\begin{abstract}
   The rapid progress of Generative Artificial Intelligence (GenAI) has enabled the effortless synthesis of high-quality visual content, while simultaneously raising pressing concerns about intellectual property protection, authenticity, and accountability. Among various countermeasures, watermarking has emerged as a fundamental mechanism for tracing provenance, distinguishing AI-generated images from natural content, and supporting trustworthy digital ecosystems. This paper presents a comprehensive survey of AI-generated image watermarking, systematically reviewing the field from five perspectives: (1) the formalization and fundamental components of image watermarking systems; (2) existing watermarking methodologies and their comparative characteristics; (3) evaluation metrics in terms of visual fidelity, embedding capacity, and detectability; (4) known vulnerabilities under malicious attacks and recent advances in secure and robust watermarking designs; and (5) open challenges, emerging trends, and future research directions. The survey seeks to offer researchers a holistic understanding of watermarking technologies for AI-generated images and to facilitate their continued advancement toward secure and responsible AI-generated content practices.
\end{abstract}



\keywords{Image watermarking, text-to-image model, AI-generated content, copyright protection, latent diffusion model, cybersecurity.}


\maketitle

\section{Introduction}
\begin{figure}[t]
    \centering
    % --------- Subfigure A: Model Evolution Timeline ---------
    \begin{subfigure}[t]{0.6\linewidth}
        \centering
        \includegraphics[width=\linewidth]{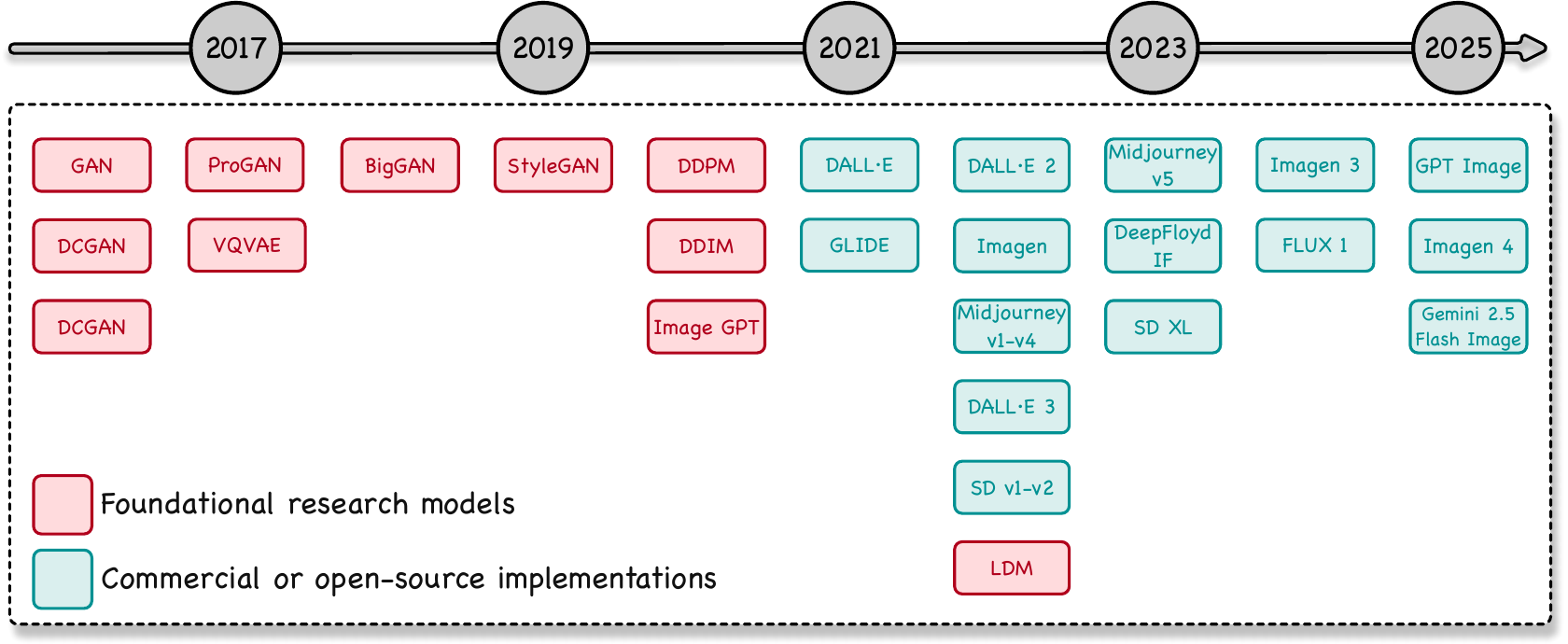}
        \caption{Evolution of generative image models, summarized from key model releases and publications from 2014 to~2025.}
        \label{fig:timeline}
    \end{subfigure}
    \hfill
    % --------- Subfigure B: Publication Trends ---------
    \begin{subfigure}[t]{0.35\linewidth}
        \centering
        \includegraphics[width=\linewidth]{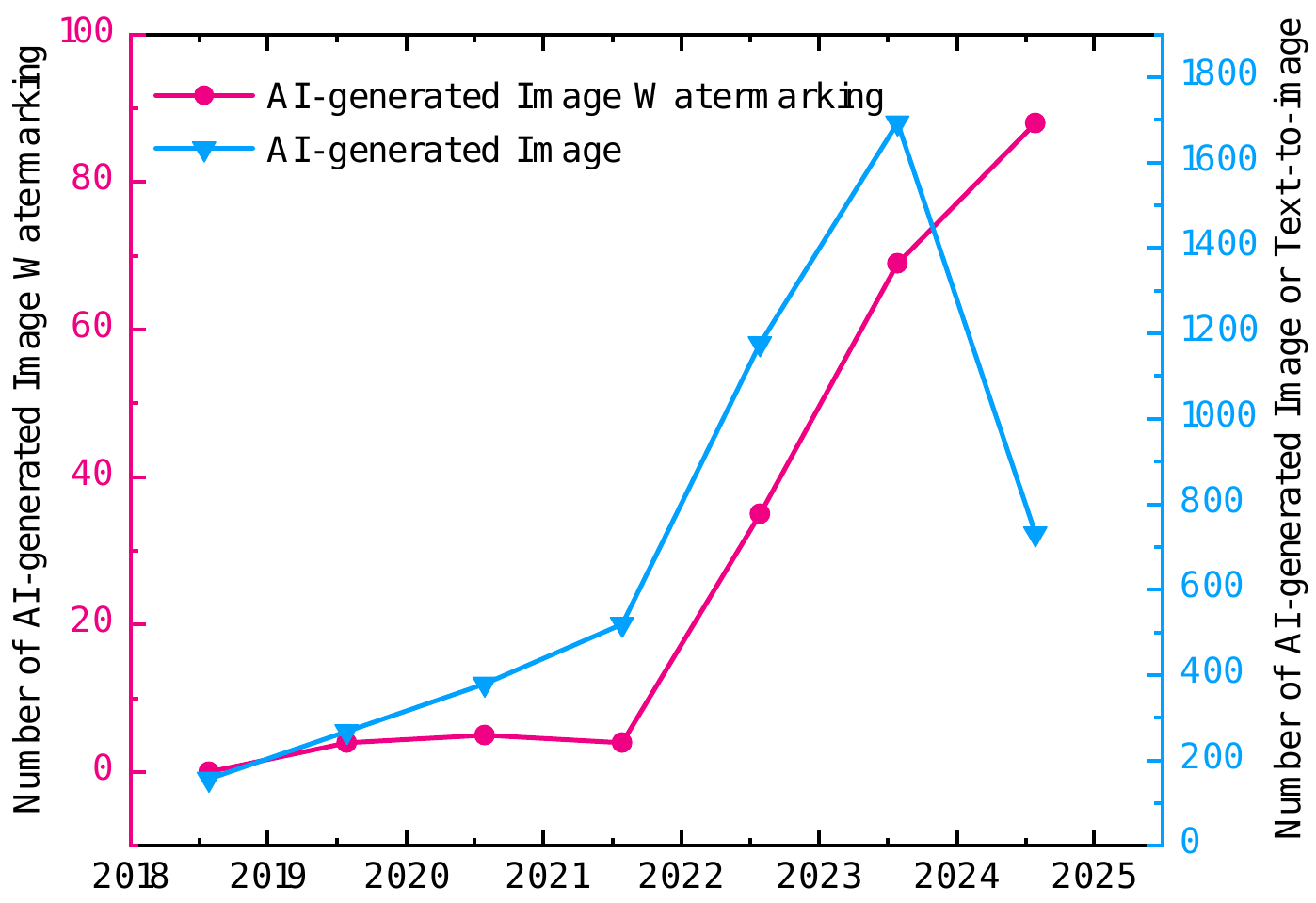}
        \caption{Publication trends in AI-generated image watermarking, based on Semantic Scholar data as of October 2025.}
        \label{fig:publications}
    \end{subfigure}
    % --------- Overall caption ---------
    \caption{Overview of the research status in AI-generated image and watermarking.}
    \label{fig:research_develop}
    \Description{This figure presents an overview of research progress in AI-generated image and watermarking, including model evolution and publication trends.}
\end{figure}

This is a burgeoning era of Generative Artificial Intelligence (GenAI), which has emerged as a transformative paradigm that focuses on creating creative content, including images~\cite{stablediffusion}, text~\cite{liangLLMSurvey}, audio~\cite{huangaudiogen}, video~\cite{sora} or even code~\cite{jiang2024survey}. In contrast to conventional discriminative models, which seek to categorize or forecast using input data, GenAI models are suggested to discover the data's underlying distribution and produce new samples using the training set~\cite{huang2025trustworthiness}. This generative ability has been made possible by advances in deep learning, particularly through architectures such as the Variational Autoencoder (VAE)~\cite{vae}, Generative Adversarial Network (GAN)~\cite{goodfellow2014generative}, and most famous diffusion model \cite{ho2020denoising,song2020denoising}, which have become the dominant backbone for contemporary GenAI systems.~\Cref{fig:timeline} presents a chronological overview of major generative model releases and publications from 2014 to 2025. 

Nowadays, GenAI has a significant impact in fields that were once thought to be exclusively human. In natural language generation, models like GPT-4~\cite{openai2023gpt4} can produce long, novel, technical explanations and conversational responses with fluency and contextual awareness like human writing. In image synthesis, commercial tools like Midjourney~\cite{Midjourney}, DALL-E~\cite{pmlr-v139-ramesh21a}, Imagen~\cite{saharia2022photorealistic} allow us to generate lifelike artistic or photorealistic images from simple textual prompts. In summary, generative AI offers exceptional efficiency and accessibility in various aspects of daily life.

However, the ease of GenAI represents a double-edged sword. While it significantly enhances creativity and productivity, it simultaneously introduces serious security, ethical, and legal concerns \cite{wu2024unveiling}. On one hand, such technologies can be exploited for malicious purposes, including the generation of deepfakes and non-consensual explicit content \cite{zhaodeepfake}. On the other hand, GenAI raises critical challenges for Intellectual Property (IP) protection, as original works created by artists and photographers are vulnerable to unauthorized modification, imitation, and reuse. This not only obscures copyright attribution but also undermines the enforcement of creators’ rights. Currently, the challenges posed by GenAI can be categorized into three main aspects:

\begin{enumerate}[noitemsep, topsep=0pt]
    \item [(1)] IP and ownership protection: In a digital environment where content is easily copied and shared, protecting IP and tracing ownership are difficult. 
    
    \item [(2)] AI-generated Content (AIGC) detection and source attribution: As the boundary between AI-generated and authentic content continues to blur, it is essential to identify not only whether the content was synthetically generated but also which model produced it. 
    
    \item [(3)] Accountability and misuse traceability: When synthetic content is exploited for harmful purposes, such as spreading false or misleading information, tracing its dissemination and identifying the responsible parties are essential yet highly challenging. 
\end{enumerate}

Addressing these three challenges introduced by generative models has become a critical research focus, particularly in the visual domain, where high-fidelity AI-generated images often blend seamlessly with human-created content. This convergence makes the questions of ownership, authenticity, and accountability increasingly difficult to resolve. Watermarking has emerged as a promising and effective approach to mitigate these pressing concerns~\cite{xu2025Advance}. As illustrated in \Cref{fig:publications}, the rapid advancement of AI-generated image technologies has simultaneously driven a flourishing research landscape  within the realm of their watermarking.

Several recent surveys have provided valuable overviews of watermarking techniques for AI-generated content, as summarized in \Cref{tab:SurveyCompare}. However, notable gaps remain that our work seeks to address. Foundational Systematization of Knowledge (SoK) studies, such as those by Zhao \textit{et al.}~\cite{zhao2024sok} and Ren \textit{et al.}~\cite{ren2024sok}, present comprehensive frameworks that formalize key definitions, properties, and threat models across multiple modalities, including text, images, and audio. Similarly, the survey by Liang \textit{et al.}~\cite{liang2024watermarking} offers an extensive review of watermarking techniques for Large Language Models (LLMs), examining cross-modal approaches and tracing their evolution from traditional methods. Nevertheless, the breadth of these works limits their ability to provide an in-depth examination of the unique security and robustness challenges inherent to the visual domain.
Other studies have focused more narrowly on AI-generated images but approach the topic from different perspectives. Luo \textit{et al.}~\cite{luo2025digital} adopt a historical viewpoint, organizing watermarking techniques from classical digital methods to contemporary diffusion-based approaches. Duan \textit{et al.}~\cite{duan2025visual} explore the interaction between diffusion models and visual watermarking, distinguishing between passive and proactive methods. However, both lack a comprehensive discussion of advanced adversarial threats and resilience under realistic attacks. In addition, in contrast to prior image-centric surveys~\cite{luo2025digital} and~\cite{duan2025visual} that primarily focus on architectural taxonomies or diffusion-based techniques, our work unifies watermarking properties, an attack taxonomy and threat model, and an evaluation protocol, thereby rethinking watermarking as a security primitive rather than a mere signal-processing tool. Overall, to address these gaps, our key contributions are summarized as follows:

\newcommand{\cmark}{\textcolor{black}{\ding{108}}} % filled circle
\newcommand{\xmark}{\textcolor{gray}{\ding{109}}} % empty circle
\newcommand{\halfblackcircle}{%

\begin{tikzpicture}[scale=0.6, baseline=-0.5ex]
    \begin{scope}
      \clip (0,0) circle (0.2);
      \fill[black] (-0.2,-0.2) rectangle (0,0.2);
      \fill[white] (0,-0.2) rectangle (0.2,0.2);
    \end{scope}
    \draw[line width=1pt] (0,0) circle (0.2);
  \end{tikzpicture}%
}

\newcommand{\blackcircle}{%
  \begin{tikzpicture}[scale=0.6, baseline=-0.5ex]
    \filldraw[black, line width=1pt] (0,0) circle (0.2);
  \end{tikzpicture}%
}

\newcommand{\whitecircle}{%
  \begin{tikzpicture}[scale=0.6, baseline=-0.5ex]
    \filldraw[white, draw=black, line width=1pt] (0,0) circle (0.2);
  \end{tikzpicture}%
}

\begin{table*}[t]
\centering
\caption{\textbf{Comparison of this survey with existing surveys on AIGC watermarking.}}

\resizebox{\textwidth}{!}{%
\begin{tabular}{
    >{\raggedright\arraybackslash}m{1.2cm}  % Paper
    >{\centering\arraybackslash}m{1.2cm}   % Year
    >{\centering\arraybackslash}m{3cm}   % Taxonomy
    >{\centering\arraybackslash}m{2.7cm}   % Core Focus
    >{\centering\arraybackslash}m{1.8cm}   % Security/Robustness
    >{\raggedright\arraybackslash}m{7cm}   % Note
}
\toprule
\textbf{Paper} & \textbf{Year} & \textbf{Taxonomy Basis} & \textbf{Modality} & \textbf{Sec./Rob. Analysis} & \textbf{Remarks} \\
\midrule
\cite{zhao2024sok} & 2024 & Property & Image, text, audio, and video & \halfblackcircle & Defines formal properties and evaluation metrics for AIGC watermarking. \\
\midrule
\cite{ren2024sok} & 2024 & Functionality and governance & Image, text, audio, and video & \whitecircle & Introduces a governance-based formulation for AIGC watermarking. \\
\midrule
\cite{luo2025digital} & 2025 & Evolutionary & Image & \halfblackcircle & Reviews the shift from traditional to AIGC watermarking. \\
\midrule
\cite{duan2025visual} & 2025 & Purpose and workflow & Image & \whitecircle & Examines diffusion–watermarking interactions and visual outcomes. \\
\midrule
\cite{liang2024watermarking} & 2025 & Modality & Image, text, audio,
multi-modal & \whitecircle & Surveys watermarking across text and image modalities in AIGC. \\
\midrule
\textbf{Ours} & 2025 & Technology-oriented & Image & \blackcircle & Provides a systematic taxonomy with joint robustness and security analysis. \\
\bottomrule
\end{tabular}
}
\begin{tablenotes}
\footnotesize
\item\textit{Sec.} and \textit{Rob.} indicate whether the survey provides security and robustness analyses of existing watermarking methods, respectively. \text{\blackcircle} denotes full support, \text{\halfblackcircle} indicates partial support, and \text{\whitecircle} means no support.
\end{tablenotes}

\label{tab:SurveyCompare}
\end{table*}

\begin{itemize}[noitemsep, topsep=3pt]
    \item  We present an extensive survey of watermarking techniques for AI-generated images. To establish a unified foundation for this emerging field, we formally define the watermarking system, decompose its fundamental components, and propose a structured taxonomy of in-generation watermarking methods. Each category is analyzed in terms of its underlying principles, technical mechanisms, strengths, and limitations.

    \item Unlike prior surveys that primarily emphasize robustness, our work treats security and robustness as equally essential pillars of trustworthy watermarking. We introduce a threat-oriented analytical framework, systematically categorize advanced attack vectors targeting AI-generated image watermarking, and review corresponding defensive strategies from a security-by-design perspective.

    \item  We consolidate evaluation methodologies across visual fidelity, embedding capacity, and statistical detectability, linking them with state-of-the-art watermarking attacks and defenses to reveal their interdependencies. Finally, we identify open challenges, discuss emerging research trends, and outline promising directions toward the design of secure, resilient, and accountable watermarking systems for responsible AIGC governance.

\end{itemize}

%\textit{Organization.} \Cref{sec:wm_role} and \Cref{sec:sec_rob_wm} examine watermarking’s role in trustworthy systems and the key challenges of ensuring robustness and security. \Cref{sec:fund_wm} establishes a unified foundation for the field and formally defines the watermarking system. \Cref{sec:trad_wm} introduces two traditional categories of watermarking. \Cref{sec:aiwm} provides background on AI-generated image watermarking and outlines two primary approaches: fine-tuning-based and initial-noise-based. \Cref{sec:evalmetric} reviews evaluation metrics for these methods, including visual quality, capacity, detectability, and existing benchmarks. \Cref{sec:attack} explores two major threats: watermark removal and watermark forgery attacks. In \Cref{sec:robustwm}, we discuss effective design strategies proposed in prior works, focusing on the security and robustness of watermarking systems. \Cref{sec:fu_work} outlines potential future research directions in this field. Finally, the survey concludes in \Cref{sec:con}.

\subsection{Watermarking and Its Role in Trustworthy AI} \label{sec:wm_role}
\begin{figure}[t]
    \centering
    \includegraphics[width=0.75\linewidth]{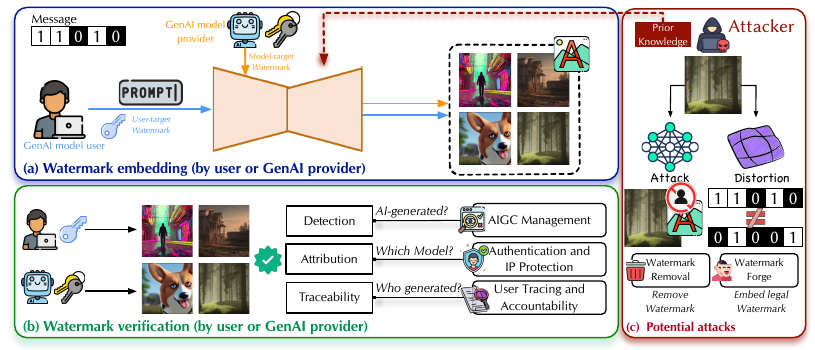}
    \caption{Watermarking scenario overview for GenAI. This figure illustrates the core components of watermarking for AI-generated images, covering watermark embedding, verification, and potential attacks.}
    \label{fig:overview}
    \Description{The figure depicts the main stages of GenAI watermarking—embedding, extraction/verification, and adversarial attacks—highlighting how watermarks are inserted, recovered, and potentially disrupted.}
\end{figure}

Digital watermarking is a technique for embedding visible or invisible identifying information into digital media to verify authenticity, assert ownership, and enable provenance tracking~\cite{zheng2007Survey,hosny2024survey}. In addition, it also serves as an effective mechanism to distinguish natural content from AIGC, thereby supporting accountability within digital ecosystems~\cite{liu2024survey,han2025robustness}. A typical digital watermarking system consists of two main stages: embedding and verification. In the embedding stage, a watermark is imperceptibly integrated into an image while preserving its perceptual quality. During verification, the embedded watermark is extracted and compared against a reference watermark or secret key to confirm authenticity or ownership. Traditional digital watermarking faces several critical limitations. First, because the watermark is applied post hoc, that is, after content generation, it is not inherently integrated into the image synthesis process, leaving it vulnerable to geometric transformations and other manipulations. Second, such methods lack semantic alignment with the generated content, as they do not exploit the latent features of the underlying generative model, thereby reducing their robustness and adaptability in the AIGC context.

In-generation watermarking for AI-generated images has emerged as a proactive and integrated strategy. This approach allows the GenAI model, user, or provider to embed watermarks or cryptographic keys during the image generation process, thereby directly addressing issues of IP protection, authenticity, and accountability. As illustrated in \Cref{fig:overview}, the watermarking mechanism subtly modifies the generated output by embedding an imperceptible yet consistent signal that serves as a verifiable marker of GenAI origin, while maintaining both semantic integrity and visual fidelity. This enables reliable post-generation detection, as the embedded watermark can be verified without dependence on statistical differences. Consequently, in-generation watermarking provides a more secure, resilient, and trustworthy means of identifying and managing AI-generated imagery.

Many countries are advancing the adoption of GenAI watermarking through emerging regulatory frameworks. China mandates watermarking of AIGC to ensure traceability and regulatory supervision~\cite{cac2025labeling}. The European Union, through the Artificial Intelligence Act~\cite{EUAIAct2024}, has established comprehensive transparency requirements, including the use of machine-readable watermarks to safeguard user rights and facilitate AIGC detection. The United States has adopted a hybrid approach, combining executive orders~\cite{whitehouse2023executive}, guidelines from the National Institute of Standards and Technology~\cite{NIST}, and voluntary industry standards to promote the implementation of watermarking technologies. Despite variations in regulatory emphasis and enforcement mechanisms, these jurisdictions converge on the recognition of GenAI watermarking as a critical instrument for combating misinformation, protecting IP, and strengthening digital trust, collectively driving its global adoption and standardization.

In parallel with regulatory efforts, major technology companies are rapidly advancing GenAI watermarking solutions. Google has introduced SynthID~\cite{synthid2025}, a proactive in-generation watermarking framework that embeds imperceptible identifiers directly during content creation. This initiative not only promotes trustworthy AI but also positions Google as a de facto standard-setter through selective open-sourcing. OpenAI, in turn, supports the C2PA standard~\cite{C2PA} and advocates for the inclusion of encrypted metadata within generated images to enable interoperable provenance verification. Meta emphasizes platform-wide transparency by integrating both visible and invisible watermarks across its services~\cite{meta}, while leveraging C2PA-compliant labels to encourage broader ecosystem adoption.

\subsection{Security and Robustness of AI-Generated Image Watermarking} \label{sec:sec_rob_wm}
Despite the growing expectations for AI-generated image watermarking, existing approaches~\cite{fernandez2023stable,wen2023tree,xiong2023flexible} remain constrained by significant limitations. Many prior methods demonstrate resilience only against basic image distortions but fail to defend against more sophisticated adversarial threats (illustrated in \Cref{fig:overview}). Recent studies~\cite{an2024waves,zhao2024invisible,wu2024robustness, wu2024robustness2,wang2024sleepermark,yang2024can,kassis2025unmarker,jain2025forging,muller2025black} have further revealed that current watermarking systems exhibit notable vulnerabilities in both robustness and security, leaving them susceptible to targeted attacks.

Security concerns the protection of watermark authenticity and resistance to forgery or misuse. Its goal is to guarantee that only authorized entities can embed and verify watermarks, thereby preventing adversaries from exploiting the mechanism for malicious purposes. Without proper security, even a robust watermark becomes unreliable; attackers could extract or replicate valid watermarks from legitimate images and reapply them to manipulated or harmful content to fabricate provenance. Such forgery attacks undermine trust, cause reputational damage, and amplify the credibility of misinformation. Secure watermarking systems, therefore, often integrate cryptographic primitives to ensure authenticity, non-repudiation, and controlled access to embedding and verification operations.

Robustness, on the other hand, pertains to a watermark’s ability to survive intentional or unintentional alterations during an image’s lifecycle. These modifications may arise from common image processing operations such as compression, cropping, resizing, or noise addition, as well as from deliberate removal attacks, where adversaries regenerate images using other generative models to erase or obscure embedded watermarks. Robust watermarking thus ensures watermark persistence and serves as the first line of defense against such attacks, safeguarding provenance and IP integrity.

In summary, these intertwined challenges of security and robustness define the central objective of our research: to design watermarking methods that ensure reliable, tamper-resistant, and verifiable provenance for AI-generated images under real-world adversarial conditions.

\section{Fundamentals of Image Watermarking}\label{sec:fund_wm}

\subsection{Notations}
We define the notations (shown in \Cref{tab:notation}) that are used in this survey.

\begin{table*}[!t]
\centering
\caption{Table of Notations}

\resizebox{\textwidth}{!}{%
\rowcolors{2}{gray!10}{white}

\begin{tabular}{l p{5cm} l p{5cm}}
\toprule
\textbf{Symbol} & \textbf{Description} & \textbf{Symbol} & \textbf{Description} \\
\midrule
\multicolumn{4}{l}{\textbf{Scalars and Dimensions}} \\
$\kappa$ & Key length. &
$n$ & Codeword length. \\
$k$ & Message length. &
$d_c$ & Carrier dimension. \\
$H,W,C$ & Image size. &
$h,w,c$ & Latent size. \\
$\tau$ & Threshold. &
$\varepsilon$ & Expected FPR. \\
$R$ & Code rate. &
$d_{\min}$ & Min. distance. \\
\midrule
\multicolumn{4}{l}{\textbf{Sets and Spaces}} \\
$\mathcal{I}$ & Image space. &
$\mathcal{Z}$ & Latent/noise space. \\
$\mathcal{C}$ & Codeword set. &
$\Theta$ & Model parameter space. \\

 $\mathcal{R}$ & Randomness space. & $\mathcal{A}$&Attack sets and adversary. \\
  $\{0,1\}^k$ & Message space. & $\{0,1\}^n$ & Codeword space.  \\ 
$\{0,1\}^\kappa$ & Cryptographic key space. && \\
\midrule
\multicolumn{4}{l}{\textbf{Variables and Parameters}} \\

$I$ & Original image. &
$I_w$ & Watermarked image. \\
$I'$ & Attacked watermarked image. &$c_K(w)$ & Keyed carrier. \\

$m$ & Message. & $w$ & Codeword. \\
$K$ & Secret key. & $\Pi$ & Public hyperparameters. \\
$y$ &  Conditioning input. &
$\theta$ & Model parameters.  \\
$\mathbf{z}_0$ & Denoised or latent representation. & $\mathbf{z}_T$ & Initial noise.  \\
$\Lambda$ & Embedding locus. &$\Delta$ &Watermark signal.  \\
$T$  & Number of steps.  &&\\% &$\gamma$ &Classifier-free guidance scale. \\
\midrule
\multicolumn{4}{l}{\textbf{Functions and Operators}} \\
$\mathsf{WM}$ & Watermarking system. &
$\mathsf{Setup}$ & Key generation. \\
$\mathsf{Encode},\mathsf{Decode}$ & Coding scheme. &
$\mathsf{Gen}$ & Generator. \\
$\mathsf{Embed}$ & Embedding function. &
$\mathsf{Extract}$ & Extraction function. \\
$\mathsf{Verify}$ & Verification function. &
$\mathsf{Channel}$ & Attack channel. \\
$S$ & Detection score. &
$d$ & Distance metric. \\
%$\mathcal{T}$ & Transform (DCT/FFT). &  $\mathcal{T}^*$ & Inverse transform. \\
$\mathcal{E}$ & Image encoder & $\mathcal{D}$ & Image decoder\\  
$\boldsymbol{\epsilon}$ & U-Net noise estimator
&$A$ & Attack function. \\
$\mathcal{W}_\text{enc}$ & Watermark encoder. & $\mathcal{W}_\text{ext}$ &Watermark extractor.\\
$\mathcal{T}$ & Text encoder. & $\phi$ & Image semantic encoder.\\
$\mathcal{L}$ & Objective function. & & \\
%\midrule
% \multicolumn{4}{l}{\textbf{Perturbations and Structures}} \\
% $\Lambda$ & Embedding locus. &
% $c_K(w)$ & Keyed carrier. \\
% $\Delta I$ & Image perturbation. &
% $\Delta \theta$ & Parameter perturbation. \\
% $\Delta \mathbf{z}_T$ & Noise perturbation. &
% $M_K$ & Keyed mask. \\
% $\Gamma_K$ & Keyed scheduler. &
% $Q_K$ & Keyed basis. \\
\bottomrule
\end{tabular}}
\label{tab:notation}
\end{table*}

% \begin{table}[h]
% \centering
% \caption{Notations}
% \begin{tabular}{clcl}
% \toprule
% \textbf{Notation} & \textbf{Description} & \textbf{Notation} & \textbf{Description} \\
% \midrule
% $\mathcal{I}$   & Image space      & $\mathcal{Z}$ & Latent space \\
% $\mathcal{C}$  &  \\
% $I$              & Cover/Input image          & $I_w$                 & Watermarked image \\
% $w$              & Watermark    & $\hat{w}$             & Extracted watermark \\
% $K$              & Cryptographic key                 \\
% $k$              & Binary key length          & $\mathcal{W}_\text{enc}$ & Watermark encoder \\
% $\mathcal{W}_\text{ext}$ & Watermark extractor & $\mathcal{L}$         & Objective function \\
% $T$              & Number of steps            & $\mathbf{z}$          & Latent representation \\
% $\mathbf{z}_T$   & Initial noise              & $\mathbf{z}_0$        & Denoised latent representation \\
% $\mathcal{E}$    & Encoder                    & $\mathcal{D}$         & Decoder \\
% $\boldsymbol{\epsilon}_\theta$ & U-Net noise estimator & $\boldsymbol{\tau}_\theta$ & Conditioning encoder \\
% $y$              & Conditioning input         & $\mathcal{F}$         & Fourier transform \\
% $\mathrm{Detect}$ & Watermark detector        & $\varepsilon$         & Pre-defined FPR value \\
% $\tau$           & Watermark detection threshold & $M$                  & Hamming distance \\
% $\delta$         & Perturbation to watermark images & $\mathcal{A}$    & Attack method \\
% \bottomrule
% \end{tabular}
% \label{tab:notation}
% \end{table}

\subsection{Terminology} \label{sec:term}
To avoid ambiguity with prior work, we define \emph{watermark} as an intentionally injected signal or structural modification embedded within the host domain. Here, we distinguish the following related concepts that are often conflated with watermarking:

\begin{description}
  \item[\textbf{Key ($K$).}] A cryptographic key used to deterministically derive security-critical elements (carriers, masks, dithers, latent subspaces, decision statistics). It is not the message, and it will never be exposed. We write $K\in\{0,1\}^{\kappa}$.
  
  \item[\textbf{Message ($m$).}] The semantic information to be conveyed or attributed (for example, model or user ID, timestamp, policy tag). We define $m\in\{0,1\}^{k}$.
  
  \item[\textbf{Codeword ($w$).}] A redundancy-added bitstring obtained by channel coding the message:
    \begin{equation}
      w=\mathsf{Encode}(m)\in\mathcal C\subseteq\{0,1\}^{n}.        
    \end{equation}

      Some papers loosely refer to $w$ as the ``watermark''; In this work, $w$ specifically denotes the encoded message bits, rather than the injected watermark signal. 

    Besides, in zero-bit watermarking ($K,m=\emptyset$) and in keyless schemes ($K=\emptyset,m\neq\emptyset$), where no explicit $\mathsf{Encode}$ is defined. For notational uniformity, we treat the $\mathsf{Encode}$ as the identity map and set:
    \begin{equation}\label{eq:id}
    \mathsf{Encode}\equiv\mathrm{id},\quad w \triangleq m,\  
    \end{equation}

  \item[\textbf{Keyed carrier $c_K(w)$.}] A continuous embedding signal obtained by modulating the codeword under the key:
  \begin{equation}
        c_K:\{0,1\}^{n}\to\mathbb R^{d_c},\quad c_K(w)=\mathsf{Modulate}_K(w),
  \end{equation} 
  where $\mathsf{Modulate}_K(w)$ maps a discrete codeword into a continuous, key-dependent watermark template. Typical realizations include spread-spectrum vectors, dithered-QIM (DC-DM) target lattice points, and latent-space directional patterns such as ring-shaped templates. Intuitively, this template specifies the ``direction'' and structure of the watermark in the chosen domain.
  \item[\textbf{Watermark ($\Delta$).}] The concrete watermark signal $\Delta$ applied to the host domain, guided by the carrier $c_K(w)$ at the injection locus $\Lambda$. Formally,
    \begin{equation}
    \begin{aligned}
          &I_w=\mathsf{Embed}_\Lambda(\underbrace{I,y,\mathbf{z}_T,\theta;\,K,w}_{\text{generate concrete}\;\Delta}), 
    \end{aligned}
    \end{equation} 
  where $I,I_w\in\mathbb{R}^{H\times W \times3 }$ denotes the clean. For instance, if $\Lambda = \mathsf{pixel}$, then $\Delta$ corresponds to a perturbation matrix of the same spatial dimensions as $I$. If $\Lambda=\mathbf{z}_T$, then $\Delta$ represents a modification applied to that latent representation.
  
  %\item[\textbf{Fingerprint.}] In this survey, a fingerprint is an active watermark assigned to a specific recipient for attribution or tracing, represented by a message $m$ encoded as $w=\mathsf{Encode}(m)$ and injected through the watermarking mechanism.
\end{description}

\subsection{Common Definition of Image Watermarking}\label{sec:comm_def}
%总体定义 （核心歧义没有定义，水印、密钥、指纹、消息这些术语究竟有何区别，适合单开一个subsection）
%Let $\mathcal I\subseteq\mathbb R^{H\times W\times C}$ be the image space and $\mathcal Z\subseteq\mathbb R^{h\times w\times c}$ be the latent/noise space for modern generators (e.g., latent diffusion model). An unmarked image is $I\in\mathcal I$, a marked image is $I_w\in\mathcal I$. 

%When AIGC generators are involved, $\mathbf{z}_T$ denotes the initial noise, and $\mathbf{z}_0$ denotes the denoised latent; $\mathcal{E},\mathcal{D}$ denote DL-based image encoder and decoder, $\mathcal{F}$ is the Fourier transform, $\varepsilon_\theta$ is a U-Net noise estimator, and $\tau_\theta$ is a conditioning encoder. A detector with a threshold $\tau$ outputs $\mathbf 1\{S(I',w;K)\ge \tau\}$ on (possibly attacked) input $I'$; it is calibrated to a target false-positive rate ($\mathrm{FPR}$) $\varepsilon$.

%Watermarks may be injected \emph{during generation} (``in-generation'', \emph{in-gen}) or \emph{after generation} into a finished image (\emph{post-hoc}). Both aim to embed an algorithmically verifiable yet visually imperceptible signal that supports attribution and provenance.

To clarify the essential components, operational flow, and security boundaries, we formally define an image watermarking system as
\begin{align}
\mathsf{WM}=(\mathsf{Setup},\mathsf{Code},\mathsf{Gen},\mathsf{Embed},\mathsf{Channel},\mathsf{Extract},\mathsf{Verify}),
\end{align}
where each component carries a clear type signature, explicit randomness, and an operational contract.

 \textbf{Setup.} A key-generation routine outputs the secret and global parameters:
\begin{equation}
\mathsf{Setup}:\;1^\kappa \mapsto (K,\Pi),\quad K\!\leftarrow\!\{0,1\}^{\kappa}.    
\end{equation}

Here, $K$ denotes the shared cryptographic key, while $\Pi$ includes public hyperparameters. The system design loosely aligns with the Kerckhoffs principle, assuming that algorithms may be public while the key $K$ remains known only to trusted embedding and verification parties.

 \textbf{Code.} A channel code provides redundancy and maps chosen messages $m$ to codewords $w$:
\begin{equation}
\begin{aligned}
&\mathsf{Code} = (\mathsf{Encode},\mathsf{Decode}),\quad \mathsf{Encode}:\{0,1\}^{k}\!\to\!\mathcal{C}\subseteq\{0,1\}^{n},\\
&w=\mathsf{Encode}(m),\quad c_k(w)=\mathsf{Modulate}_K(w).
\end{aligned}
\end{equation}

%Zero-bit watermarking sets $k=0$ (presence detection only). No explicit message $m$ is defined, and $\mathsf{Verify}$ operates directly on the codeword $\hat{w}$. 
For watermarking with \(k > 0\), let \(R = k/n\) and \(d_{\min}\) denote the code rate and minimum distance of \(\mathcal{C}\), respectively. Under hard decoding, up to \(\lfloor(d_{\min}-1)/2\rfloor\) errors are correctable. Then, a keyed modulation map \(c_K:\{0,1\}^n \to \mathbb{R}^{d_c}\) lifts codeword \(w\) to a continuous carrier \(c_K(w)\).

 \textbf{Gen.} An image generator (the identity operator or image reconstructor in the post-hoc watermarking system) produces the image:
\begin{equation}
\mathsf{Gen}: \Theta\times\mathcal Z\times\mathcal Y \to \mathcal I,\quad 
I=\mathsf{Gen}(\theta,\mathbf{z}_T,y),    
\end{equation}  
where \(y\in\mathcal Y\) denotes the conditioning input; if unused, we take \(y=\varnothing\).

%Some in-generation watermarking additionally expose an approximate inversion $\mathsf{Inv}_\theta:\mathcal I\times\mathcal Y\to\mathcal Z$ (e.g., DDIM inversion \cite{song2020denoising}) used for verification.

 \textbf{Embed.}
%An embedding operator parameterized by $\Lambda$ (defined in~\S\ref{subsec:embed}):
% \[
% I_w=\mathsf{Embed}_\Lambda(I,y,\mathbf{z}_T,\theta;K,w).
% \]
The locus component of $\Lambda$ selects where the mark is injected
\begin{equation}
\begin{aligned}
\mathsf{Embed}:\;&
\mathcal I\times\mathcal Y\times\mathcal Z\times\Theta\times
\{0,1\}^{\kappa}\times\{0,1\}^{n}\times\mathcal R \;\to\; \mathcal I,\\
&I_w \;=\; \mathsf{Embed}_\Lambda\big( I, y, \mathbf z_T, \theta;K, c_K(w) ,\rho \big),
\end{aligned}
\end{equation}
where $\rho\in\mathcal R$ denotes a nonce used by the embedder (e.g., randomized masks/schedulers, or sampling noise not captured by $\mathbf z_T$). It is independent of the key $K$ and may be public side information; if embedding is deterministic given \((I,y,\mathbf z_T,\theta;K,c_k(w))\), we take $\rho=\bot$. $c_K(w)$ will determine the concrete watermark signal $\Delta$ to be embedded into the host domain. 

The locus $\Lambda$ selects \emph{where} the watermark is injected:
\begin{equation}
\textsf{locus}\in\{\textsf{pixel},\textsf{freq},\textsf{feature map},\,\mathbf{z}_T,\,\theta\}.    
\end{equation}

Besides, the embedding specifies how to inject, e.g., an additive update, modulation/quantization, masking, prompt gating, or distribution-preserving sampling. The embedding must satisfy: (i) image quality; and (ii) detectability.

 \textbf{Channel.} A (possibly adversarial) family of image transformations:
\begin{equation}
    \mathsf{Channel}:\;\mathcal A\subseteq\{A:\mathcal I\to\mathcal I\},\quad 
I' = A(I_w;\rho),
\end{equation}
including benign image processing and malicious attack. 
%Analyses distinguish a benign distribution $A\!\sim\!\mathsf{Benign}$ from an adversarial choice $A\!\in\!\mathsf{Adv}$.

 \textbf{Extract.}
The extractor returns both the recovered codeword and the decoded message (or $\perp$ on failure):
\begin{equation}
\begin{aligned}
&\mathsf{Extract}:\;\mathcal I\times\{0,1\}^{\kappa}\times\Theta\times\mathcal Y
\to \big(\{0,1\}^{n}\times\{0,1\}^{k}\big)\cup\{\perp\},\\
&\hat w \;=\; \mathsf{Extract}(I';K,\theta,y), \quad \hat m \;=\; \mathsf{Decode}(\hat w),\quad \mathsf{Decode}\equiv\mathsf{id}\text{ if }K=\emptyset.
\end{aligned}
\end{equation}

By \Cref{eq:id}, for $K=\emptyset$ and there is no explicit (\(\mathsf{Encode}\)) is defined, thus, the $\mathsf{Decode}$ is also an identity map and set, so that the subsequent formula can be written in the same form. Verification, therefore, operates directly on the extracted message \(\hat m\). 
  
%For initial-noise schemes, extraction may begin with $\tilde {\mathbf{z}}_T=\mathsf{Inv}_\theta(I',y)$ and then apply a latent-space statistic.

 \textbf{Verify.} A statistical test maps the extractor output or a score to a binary verdict:
\begin{equation}
\mathsf{Verify}:\;\big(\hat m,m\big)\mapsto 
\mathbf 1\{S(\hat m,m)\ge \tau\},
\end{equation}
where \(S\) denotes a detection score computed from the decoded message $\hat m$ and original $m$. By convention, $S$ measures the similarity between $m$ and $\hat m$, and larger \(S\) implies stronger evidence of verification; if \(S\) is a distance, the decision inequality is reversed (smaller is better). \(\tau\) is the score threshold, calibrated to meet a target False Positive Rate ($\mathrm{FPR}$) \(\varepsilon\) on unmarked data. Practical details will be discussed in \Cref{sec:dete_metric}.

% The operational metric is $\mathrm{FPR}=\varepsilon$. In multi-user attribution, the global error is $\mathrm{FPR}(\tau,N)=1-(1-\mathrm{FPR}(\tau))^{N}$ for $N$ identities.
% Zero-bit (presence-only) variant: replace $S(I',w;K)$ with $S(I';K)$.
%暂时将密钥k跟水印w同时放进来，明日找曹捷确认已有水印都是如何

\medskip
% \textbf{Unified pipeline.}
Overall, the whole pipeline of image watermarking can be shown as follows:
% \[
% m \xrightarrow{\ \mathsf{Encode}\ }\ w \xrightarrow{\ \mathsf{Mod}_K\ }\ c_K(w)
% \ \xrightarrow{\ \mathsf{Embed}_\Lambda(\cdot;K)\ }\ I_w
% \ \xrightarrow{\ A\in\mathcal A\ }\ I'
% \ \xrightarrow{\ \mathsf{Verify}(\cdot;K)\ }\ \mathbf{ 1/0}
% \]
\begin{equation}
\begin{aligned}
&m \xrightarrow{\ \mathsf{Encode}\ } w
   \xrightarrow{\ \mathsf{Modulate}_K\ } c_K(w) \xrightarrow{\ \mathsf{Embed}_\Lambda\big( I, y, \mathbf z_T, \theta ;K, w ,\rho \big)\ }  I_w\xrightarrow{\ A\in\mathcal A\ } I'\xrightarrow{\mathsf{Extract}(I';K,\theta)} \\
&\hat m 
   \xrightarrow{\ \mathsf{Verify}(\hat m, m)\ } \{\mathit{false},\mathit{true}\}.
\end{aligned}
\end{equation}

\subsection{Image Watermarking Properties} \label{sec:wm_prop}
\subsubsection{Quality}
Quality is a fundamental requirement for a watermarking system. It comprises two independent and equally critical dimensions: the image quality and the extraction quality of the embedded message.

Image quality refers to the visual and semantic fidelity of the watermarked image. The watermark signal should be imperceptible, introducing no perceivable artifacts or semantic deviations.
\begin{myDef}
A watermarking system is said to achieve high image quality at thresholds $(\delta,\eta)$ if, for any key $K$ and message $m$, the watermarked image $I_w$ satisfies
\begin{equation}
d_{\mathrm{vis}}(I, I_w) \le \delta
\quad\text{and}\quad
d_{\mathrm{sem}}\!\big(y, \phi(I_w)\big) \le \eta,
\end{equation}
where $d_{\mathrm{vis}}$ denotes perceptual distance metric and $d_{\mathrm{sem}}$ is the semantic alignment between $y$ and $I_w$ via image semantic encoder $\phi$. The constants $\delta$ and $\eta$ are small, application-dependent.
\end{myDef}

Message quality refers to the ability of the extractor $\mathsf{Extract}$ to recover the codeword $w$ from the $I_w$ and then decode the original message $m$ accurately, in the absence of distortions or attacks. 

\begin{myDef}
A watermarking system achieves high message quality if, on unattacked watermarked images, the authorized detector with the legal key can verify the watermark and recover the original message with negligible failure:
\begin{equation}
\begin{aligned}
\Pr\big[\, \mathsf{Verify}(\hat m,m)\rightarrow \mathit{true} : \hat m \gets \mathsf{Extract}(I_w; K, \theta) \,\big] \ge\; 1-\mathrm{negl}(\lambda),
\end{aligned}
\end{equation}
\end{myDef}
where $\lambda$ denotes the security parameter.

\subsubsection{Robustness} \label{sec:rob_define}
Robustness is one of the core attributes for evaluating the effectiveness of watermarking techniques. It refers to the ability of the embedded watermark signal to remain detectable or verifiable by the detector even after the image has undergone various intentional or unintentional transformations, commonly referred to as ``attack''.

\begin{myDef}
A watermarking method is said to be robust if, for some attacks 
$A\in \mathcal{A}$ applied to a watermarked image $I_w$, the probability that the detector fails to recognize the watermark is bounded by a function $\mathsf{negl}(\lambda)$:

\begin{equation}
\begin{aligned}
    \Pr\big[\mathsf{Verify}(\hat m, m) \to \mathit{false}  : \hat m \leftarrow \mathsf{Extract}(I';K,\theta)\big] \leq \mathsf{negl}(\lambda),
\end{aligned}
\end{equation}
where $I'$ denotes the watermarked image that has been attacked through $\mathsf{Channel}$. 
\end{myDef}

Robustness does not imply resistance against all possible attacks. Specifically, the set $\mathcal{A}$ does not include all conceivable manipulations. Robustness is defined against a pre-specified class of attacks (e.g., compression, noise, scaling) that preserve the image's perceptual utility. It does not guarantee resistance to attacks designed to completely destroy the data. Besides, the definition is a probabilistic security guarantee. It allows for a negligibly small probability of failure $\mathsf{negl}(\lambda)$, acknowledging that even a valid attack $A \in \mathcal{A}$ (e.g., extreme compression) might statistically overwhelm the watermark signal.

\subsubsection{Security}
Security in watermarking is intended to withstand forgery attacks and unauthorized watermark extraction. The adversary’s objectives typically involve impersonation, information theft, or undermining the credibility of the system. The foundation of security lies in the secret key $K$, which is exclusively known to the authorized embedding and verification parties. Based on this principle, the security of a watermarking system is generally characterized by two essential properties: unforgeability and imperceptibility.

\paragraph{Unforgeability.} Unforgeability requires that without access to the secret key $K$, an adversary cannot fabricate a forged watermarked image $I'$ that will be accepted as valid by $\mathsf{Verify}$. In other words, it should be computationally infeasible to produce an image that appears to carry a legitimate watermark without going through the authorized embedding process. 

\begin{myDef}
A watermarking scheme is said to be unforgeable if, for any probabilistic polynomial-time adversary $\mathcal{A}$ without knowledge of the secret key $K$, the probability of successfully creating a valid watermark is negligible:
\begin{equation}
\begin{aligned}
   \Pr \big[ \mathsf{Verify}(\hat m, m) \rightarrow \mathit{true} :    \hat m\leftarrow\mathsf{Extract}(I';K,\theta), I' 
   \leftarrow A(I, 1^\lambda) \big] 
    \leq \text{negl}(\lambda),\;\text{for}\;A\in\mathcal{A}.
    \end{aligned}
\end{equation}
\end{myDef}

\paragraph{Imperceptibility.} It is also referred to as covertness or computational security, which requires that, in the absence of the secret key $K$, an adversary cannot distinguish a watermarked image $I_w$ from an unwatermarked image $I$. In other words, the watermark signal should be computationally indistinguishable from the natural noise inherent in the image. This can be formalized by an indistinguishability game.

\begin{myDef}
A watermarking scheme is said to be imperceptible if, for any probabilistic polynomial-time adversary (distinguisher) $\mathcal{A}$ without knowledge of $K$, the advantage in distinguishing between a watermarked image and a non-watermarked image is negligible:
\begin{equation}
   \big| \Pr[\mathcal{A}(I_w)\rightarrow \mathit{true}] - \Pr[\mathcal{A}(I)\rightarrow \mathit{true}] \big| \leq \text{negl}(\lambda).
\end{equation}
\end{myDef}
This definition ensures that the very existence of the watermark remains concealed, an essential property in real scenarios, while simultaneously preventing adversaries from mounting targeted attacks based on watermark detection.

In short, robustness concerns survivability under bounded distortions, whereas
security provides cryptographic indistinguishability and unforgeability
guarantees under secret keys.

\section{Traditional Image Watermarking}\label{sec:trad_wm}
Image watermarking has long been a well-established area of research, with a rich history spanning both theoretical development and practical deployment \cite{al2007combined,hosny2024survey}. Traditional image watermarking techniques can be broadly categorized into two main classes: domain-based methods and deep learning-based methods. In this section, we provide a brief overview of these two lines of research, outlining their foundational principles, representative techniques, and limitations.

\subsection{Domain-based Image Watermarking}\label{sec:domain_wm}

Traditional digital image watermarking methods are characterized by their reliance on hand-crafted watermark embedding and extraction algorithms primarily rooted in signal processing principles. We formalize domain-based image watermarking using the following pipeline based on \Cref{sec:comm_def}.

\begin{equation}\label{eq:tradi_pipe}
\begin{aligned}
&m \xrightarrow{\ \mathsf{Encode}\ } w
    \xrightarrow{\ \mathsf{Modulate}_K\ } c_K(w) \xrightarrow{\ \mathsf{Embed}_\Lambda\in\{\mathsf{pixel,trans}\}\big( I, y=\emptyset, \mathbf z_T=\emptyset, \theta=\emptyset; K,w,\rho \big)\ } I_w \xrightarrow{\ A\in\mathcal A\ } \\
    &I'\xrightarrow{\mathsf{Extract}(I'; K, \theta=\emptyset)} \hat m \xrightarrow{\ \mathsf{Verify}(\hat m, m)\ } \{\mathit{false},\mathit{true}\}.
\end{aligned}
\end{equation}

\begin{figure}[ht]
    \centering
    \includegraphics[width=0.6\linewidth]{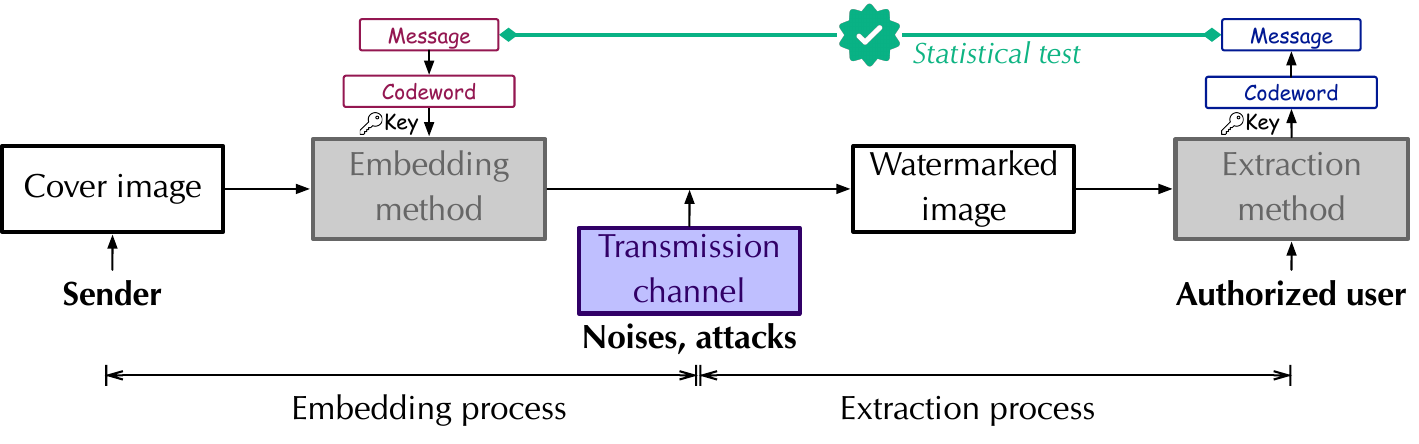}
    \caption{Traditional image watermarking pipeline.}
    \label{fig:traditional image watermark}
    \Description{A flowchart of a traditional invisible image watermarking pipeline. The original image is passed through a watermark embedder to produce a watermarked image. An optional attack channel may distort the image before it is passed to a decoder for message recovery.}
\end{figure}

This process begins with an existing cover image I, where the embedding locus $\Lambda$ is typically constrained by $\mathsf{Embed}$ to lie within specific host domains of either the spatial or the transform domain. The message $m$ is first encoded into the codeword $w$, which is then modulated under a secret key $K$ and injected as a watermark signal into $I$ to produce the watermarked image $I_w$. After passing through a potential attack channel $\mathcal{A}$, an authorized user holding $K$ applies $\mathsf{Extract}$ to recover $\hat m$, and $\mathsf{Verify}$ compares $\hat m$ with the original $m$ to decide authenticity, as shown in \Cref{fig:traditional image watermark}. The $\mathsf{Encode}$, $\mathsf{Modulate}_{K}$, and $\mathsf{Embed}$ functions are usually integrated and jointly designed as a single embedding method. Domain-based methods can be categorized into the following two categories based on the embedding locus $\Lambda$~\cite{wan2022comprehensive}.

\textbf{Spatial Domain Methods}: These methods directly alter the pixel values of an image to contain the message. For example, Schyndel \textit{et al.}~\cite{van1994digital} proposed the Least Significant Bit (LSB) method, which embeds a message into the least significant bits of image pixels, since the least significant bit has minimal impact on visual quality. However, this LSB method is vulnerable to common image processing operations. To address this, Lee \textit{et al.}~\cite{lee2008new} introduced a randomized embedding technique using a seeded mapping function to reduce the risk of extraction by eavesdroppers, demonstrating improved robustness against cropping and histogram equalization attacks.

\textbf{Transform Domain Methods}: These approaches transform an image into a different mathematical space, by Discrete Cosine Transform (DCT)~\cite{al2007combined} or Discrete Wavelet Transform (DWT), and embed the message by altering the parameter's value in this transformed domain. Methods like DWT-DCT~\cite{al2007combined} and DWT-DCT-SVD~\cite{navas2008dwt} are specific examples. Embedding in the transform domain often allows the message to be spread across the image, potentially offering more robustness against manipulations compared to the spatial methods.

\subsection{Deep Learning-based Image Watermarking}\label{sec:dlwm}

Deep learning (DL)-based image watermarking leverages Deep Neural Networks (DNNs) to embed messages into cover images. Similar to \Cref{eq:tradi_pipe}, the formal pipeline of DL-based methods differs from traditional domain-based schemes in that the embedding locus $\Lambda=\mathsf{\{feature\;map,\theta\}}$, and $\theta \neq \varnothing$, since these methods rely on one or more DNNs trained to reconstruct the watermarked image so that it closely approximates the original. Such networks can be regarded as a specialized instantiation of $\mathsf{Gen}$.

Kandi \textit{et al.}~\cite{kandi2017exploring} was the first to explore the potential of Convolutional Neural Network (CNN)-based encoder–decoder architecture for digital image watermarking. Zhu \textit{et al.}~\cite{zhu2018hidden} proposed HiDDeN, a pioneering work targeting DL-based invisible image watermarking. They jointly trained a CNN-based autoencoder network. The encoder takes a message $m$ and a cover image $I$ as input and generates a watermarked image visually indistinguishable from the cover image. The image watermark extractor then accurately recovers the embedded message. Specifically, the objective function of this method is defined as:
\begin{equation}\label{eq:hidden_loss}
\begin{aligned}
    &\mathcal{L}(\theta) = \underbrace{\mathcal{L}_\text{BSE}\left(m, \mathcal{W}_\text{ext}(I_w)\right)}_{\text{Recovery loss}}\quad + \underbrace{\alpha \mathcal{L}_\text{IMG}\left(I, I_w\right)}_{\text{Image reconstruction loss}},\\
    &\textit{where}\; I_w = \mathcal{D}(\mathcal{E}(I)||\mathcal{W}_\text{enc}(m)),\;\theta = \bigl(
    \theta_{\mathcal{E}},
    \theta_{\mathcal{D}},
    \theta_{\mathcal{W}_\text{enc}},
    \theta_{\mathcal{W}_\text{ext}}\bigr).\\
\end{aligned}
\end{equation}

\begin{figure}[t]
    \centering
    \includegraphics[width=0.6\linewidth]{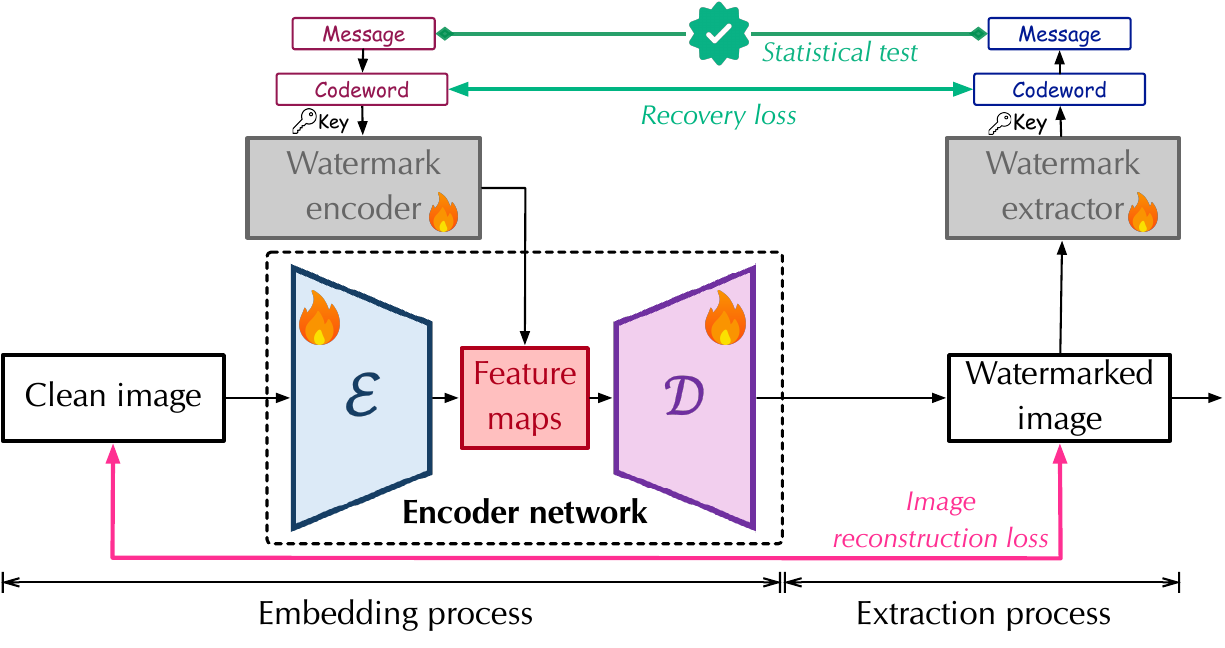}
    \caption{DL-based image watermarking method architecture.}
    \Description{A schematic diagram showing the architecture of a deep learning-based image watermarking system.}
    \label{fig:DL_wm}
\end{figure}

As shown in \Cref{fig:DL_wm}, the DL-based watermarking architecture commonly comprises four main components: an image encoder $\mathcal{E}$, an image decoder $\mathcal{D}$, a watermark encoder $\mathcal{W}_{\text{enc}}$, and a watermark extractor $\mathcal{W}_\text{ext}$. The watermark encoder can be viewed as integrating the operations 
$\mathcal{W}_{\text{enc}} = \{\mathsf{Encode}, \mathsf{Modulate}_K, \mathsf{Embed}\}$, while the extractor corresponds to $\mathcal{W}_{\text{ext}} = \{\mathsf{Extract}\}$.The model is jointly optimized via binary cross-entropy loss $\mathcal{L}_\text{BSE}$ and image reconstruction loss $\mathcal{L}_\text{IMG}$, typically using Mean Squared Error (MSE) loss.

Building on HiDDeN, StegaStamp~\cite{tancik2020stegastamp} specifically targets the challenging scenario of physical transmission by introducing differentiable image perturbations that closely approximate distortions arising from real-world photography. RivaGAN, proposed by Zhang et al.~\cite{zhang2019robust}, adopts an encoder–decoder structure with an attention mechanism for content-aware bit embedding. A dual-adversary framework jointly optimizes visual fidelity and resistance to watermark removal. Beyond these works, a range of innovative training strategies and model designs have emerged to further improve watermarking robustness. For instance, Luo \textit{et al.}~\cite{luo2020distortion} leverage adversarial training to counteract the detrimental effects of unknown distortions on embedded messages. Similarly, self-supervised learning has been employed to embed marks and binary information directly into the latent spaces of pre-trained self-supervised models such as DINO~\cite{caron2021emerging}, demonstrating the potential of latent-space representations for enhancing quality.

\section{Watermarking for AI-Generated Images}\label{sec:aiwm}
Image generation technologies have advanced dramatically in recent years. A wide range of visual generative models, such as variational autoencoders (VAEs)~\cite{vae}, flow-based models~\cite{kingma2018glow}, generative adversarial networks (GANs)~\cite{goodfellow2014generative,karras2019style}, and diffusion models~\cite{ho2020denoising,song2020denoising,rombach2022high}, have been introduced. Among these, diffusion models have emerged as the dominant paradigm in modern image synthesis, powering commercial tools such as DALL·E~\cite{ramesh2021zero} and Midjourney~\cite{Midjourney}. These systems allow users with little to no technical expertise to generate photorealistic images and perform sophisticated content editing with ease. While these advances have significantly enhanced creativity and productivity, they have also introduced serious challenges related to authenticity, ownership, and misuse. Consequently, effective watermarking techniques for AI-generated images are urgently needed to promote a safe and trustworthy digital ecosystem.

Traditional image watermarking methods, which operate in a post-hoc manner, can indeed be applied to AI-generated images after generation. The recent literature contains only a small number of post-hoc methods for watermarking AI-generated images (e.g., \cite{xian2024raw,lu2024robust}). Because they suffer from several fundamental limitations: (i) post-hoc watermarking, not being integrated into image synthesis, is highly susceptible to distortions and adversarial attacks, compromising robustness; (ii) it also lacks semantic alignment, as it ignores the model’s internal representations, limiting its utility for provenance, ownership, and accountability.

\begin{figure}[t]
    \centering
    \includegraphics[width=0.7\linewidth]{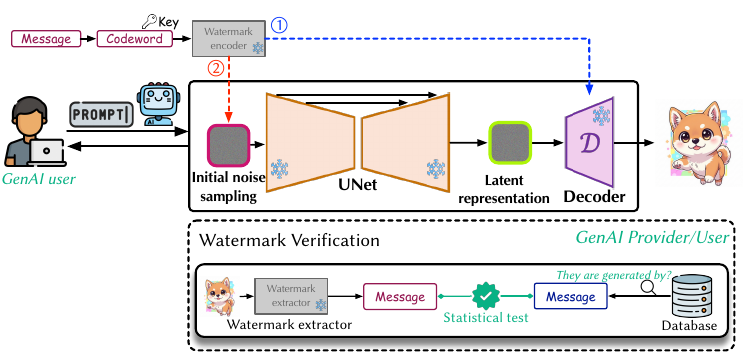}
    \caption{AI-generated image watermarking overview. \ding{192} fine-tuning-based watermarking; \ding{193} initial noise-based watermarking.}
    \label{fig:aigiwatermark}
    \Description{Two-part diagram showing AI-generated image watermarking methods. The first part illustrates fine-tuning-based watermarking, where the generative model is adapted to embed watermarks during generation. The second part shows initial noise-based watermarking, where the watermark is injected into the initial noise input, leaving the model unchanged.}
\end{figure}

To address these limitations, growing attention has been directed toward in-generation watermarking methods, which embed the watermark directly within the generative process itself. This approach enhances resilience, interpretability, and traceability, as the watermark becomes an intrinsic component of image synthesis rather than an external addition. In-generation watermarking techniques for AI-generated images can be categorized according to where and how the watermark is embedded during generation. As illustrated in \Cref{fig:aigiwatermark}, two mainstream paradigms have emerged: fine-tuning-based watermarking and initial-noise-based watermarking. In the following subsections, we first outline the widely adopted Latent Diffusion Model (LDM) image generation pipeline to provide the necessary background. We then present a comparative discussion of these two watermarking approaches, highlighting their key mechanisms, design philosophies, and representative implementations. 

\subsection{Stakeholders in AI-Generated Image Watermarking}
In the AI-generated image watermarking system, several key stakeholders interact, each defined by distinct objectives, capabilities, and motivations. Based on the system's design, these roles can be primarily categorized as the AI model provider, the model user, and the potential attacker.

\paragraph{GenAI Provider} It is the entity responsible for developing and deploying the generative models. The provider's main objective is the protection of their model's output. To achieve this, providers typically employ \textit{``model-targeted watermarking'}'. This approach involves embedding a unique, imperceptible identifier that represents the specific model into all generated outputs. As illustrated in \Cref{fig:aigiwatermark}, the provider manages this embedding process and typically ``retains ownership of the watermark extractor''. This centralized control allows them to reliably verify if a suspect image was synthesized by their model, thus enabling content attribution and IP protection.

\paragraph{GenAI User} It is the individual or entity who utilizes the provider's model to create content. The user's goal is to enable ``content tracing and personal IP protection''. They seek to protect their own creations from unauthorized copying or redistribution and to be able to trace their content if it is misused. This user-centric need is met by \textit{``user-targeted watermarking''}. In this content, the system assigns a unique message, such as a user ID, to each specific user. This message is then embedded into all content that the user generates. This allows the provider or system administrator to ``attribute the generated image to the user'' which is essential for accountability and for tracing the source of leaked or misused content. Moreover, the model user may also control the watermarking system (without relying on the provider) to watermark their own outputs. This allows the user to assert authorship and provenance for IP protection.

\paragraph{Potential Adversary.} It is a malicious party whose objective is to compromise the integrity, security, or reliability of the watermarking system. An adversary's efforts focus mainly on two tasks: watermark removal or watermark forgery. Watermark removal attacks seek to erase embedded watermarks while preserving image quality, using techniques like regeneration, adversarial perturbations, or model fine-tuning. In contrast, forgery attacks aim to embed valid watermarks into unrelated content to falsify origin or ownership.

\subsection{Diffusion Model}\label{sec:dm}
Diffusion-based techniques have rapidly become the leading paradigm in image generation, demonstrating remarkable performance in synthesizing high-quality and diverse visual content. These models are inspired by principles of nonequilibrium thermodynamics, where data generation is conceptualized as a gradual denoising process that reverses a forward diffusion of noise. The foundational work by Sohl-Dickstein \textit{et al.}~\cite{sohl2015deep} introduced diffusion probabilistic models, which progressively add Gaussian noise to data and subsequently learn to reverse this process. However, early diffusion models received limited attention due to computational inefficiency and suboptimal generation quality, lagging behind alternatives such as GANs and VAEs. A major breakthrough arrived with the Denoising Diffusion Probabilistic Model (DDPM)~\cite{ho2020denoising}, which greatly improved both sample fidelity and training stability. Building upon DDPM, Song \textit{et al.}~\cite{song2020denoising} proposed the Denoising Diffusion Implicit Model (DDIM), introducing a deterministic sampling procedure that accelerates inference and enables image inversion, the recovery of latent noise from real images. Subsequently, Rombach \textit{et al.}~\cite{rombach2022high} developed the Latent Diffusion Model (LDM), which performs diffusion in latent feature space rather than the pixel space, achieving high-resolution synthesis with practical computational efficiency. LDM has since become the backbone of many widely used systems, including Stable Diffusion~\cite{stablediffusion} and Midjourney~\cite{Midjourney}, firmly establishing diffusion models as the cornerstone of today’s generative image ecosystem.

LDM performs the diffusion process not directly in the high-dimensional image space $\mathcal{I}$, but rather in a more compact latent space $\mathcal{Z}$. A VAE is first trained to learn an encoder $\mathcal{E}$ and a decoder $\mathcal{D}$, such that an image $I$ can be mapped into a latent representation $\mathbf{z}_0 = \mathcal{E}(I)$ and approximately reconstructed as $\hat{I} = \mathcal{D}(\mathbf{z}_0)$. Once trained, the VAE parameters are frozen. The denoising model based on U-Net is then trained in this latent space: the forward process gradually   $\mathbf{z}_0$ into a noisy latent $\mathbf{z}_T$, and the reverse denoising process reconstructs a clean latent $\hat{\mathbf{z}}_0$. Finally, the decoder $\mathcal{D}$ maps $\hat{\mathbf{z}}_0$ back into the image space to obtain the generated image $\tilde{I} = \mathcal{D}(\hat{\mathbf{z}}_0)$.

The forward diffusion process is a Markov chain that gradually adds Gaussian noise to the latent representation $\mathbf{z}_0$ until it becomes pure noise $\mathbf{z}_T$. For the reverse denoising process, DDIM provides an efficient deterministic sampling strategy, mapping from $\mathbf{z}_T$ to $\mathbf{z}_0$. For each denoising step, a noise-predictor $\epsilon_\theta$ estimates the noise $\epsilon_\theta(\mathbf{z}_t)$ added to $\mathbf{z}_0$. LDM adopts a training objective, as defined in~\Cref{eq:ldm}, minimizing the distance between the predicted noise $\epsilon_\theta(\mathbf{z}_t, t)$ and the ground-truth noise $\epsilon$. 
\begin{equation}
\mathcal{L}_{\text{LDM}}=
\mathbb{E}_{\substack{\mathcal{E}(I), t\sim\mathcal{U}(0,T-1),\epsilon\sim\mathcal{N}(0,\mathbf{I})}}
\Bigl[\bigl\lVert \epsilon-\boldsymbol{\epsilon}_{\theta}(z_t,t,\mathcal{T}_{\theta}(y))\bigr\rVert_2^2\Bigr],
\label{eq:ldm}
\end{equation}
where $\boldsymbol{\epsilon}$ denotes U-Net~\cite{ronneberger2015u} denoiser. To enable controllable generation, LDM incorporates a text input $y$ using a text encoder $\mathcal{T}$ based on BERT~\cite{devlin2019bert} or CLIP~\cite{radford2021learning}.

SDM~\cite{rombach2022high} offers a practical, open-source implementation of the LDM framework. It leverages a frozen VAE for encoding/decoding and trains a U-Net denoiser in latent space, enabling efficient, scalable generation. Due to its quality, modularity, and accessibility, SDM is the standard backbone for recent AI image watermarking research.

\definecolor{main}{RGB}{191,148,228}
\definecolor{sub}{RGB}{223,199,247}
\definecolor{forgerymain}{RGB}{0,191,255}
\definecolor{forgerysub}{RGB}{180,235,255}
\definecolor{advgold}{RGB}{218,145,0}
\definecolor{advlight}{RGB}{245,210,130}
\definecolor{key}{RGB}{200,210,100}

\definecolor{regenred}{RGB}{240,128,128}
\definecolor{regenlight}{RGB}{255,200,200}
\definecolor{citegray}{gray}{0.95}

\begin{figure}[t]
\centering
\tikzset{
  my node/.style={
    draw=black, semithick, align=center, text width=1.2cm, rounded corners=3,
  },
  my leaf/.style={
    draw=black, semithick, align=left, text width=8.5cm, rounded corners=3,
  }
}
\forestset{
  every leaf node/.style={if n children=0{#1}{}},
  every tree node/.style={if n children=0{minimum width=1em}{#1}},
}

\scalebox{0.85}{%
\begin{forest}
  nonleaf/.style={font=\scriptsize},
  for tree={
    every leaf node={my leaf, font=\tiny},
    every tree node={my node, font=\tiny, l sep-=4.5pt, l-=1.pt},
    anchor=west, inner sep=2pt,
    l sep=10pt, s sep=3pt, fit=tight,
    grow'=east, edge={semithick},
    parent anchor=east, child anchor=west,
    if n children=0{}{nonleaf},
    edge path={\noexpand\path[draw,\forestoption{edge}]
      (!u.parent anchor) -- +(5pt,0) |- (.child anchor)\forestoption{edge label};},
    if={isodd(n_children())}{
      for children={
        if={equal(n,(n_children("!u")+1)/2)}{calign with current}{}
      }
    }{}
  }
  [{\textbf{AI-generated Image\\ Watermarking}}, draw=black, semithick, fill=gray!15, text width=3cm, text=black
    [{\textbf{Fine-tuning-based Watermarking\\(\cref{sec:fine-tune_wm})}}, draw=black, semithick, fill=main!25, text width=2.5cm, text=black
      [{\footnotesize~\cite{fernandez2023stable},~\cite{xiong2023flexible},~\cite{kim2024wouaf},~\cite{wang2024sleepermark},~\cite{ciwmadapter},~\cite{feng2024aqualora},~\cite{lu2024robust}}, draw=gray!50, fill=citegray, text=black,text width=4.5cm]
    ]
    [{\textbf{Initial-noise-based Watermarking\\(\cref{sec:ini_wm})}}, draw=black, semithick, fill=forgerymain!25, text width=2.5cm, text=black
      [{Direct embedding\\(\cref{sec:dir_emb})}, draw=black, semithick, fill=regenlight, text width=2.5cm, text=black
        [{\footnotesize~\cite{wen2023tree},~\cite{zhu2024flexible},~\cite{ci2024ringid}}, draw=gray!50, fill=citegray, text=black,text width=2cm]
      ]
      [{Iterative embedding\\(\cref{sec:iter_emb})}, draw=black, semithick, fill=advlight, text width=2.5cm, text=black
        [{\footnotesize~\cite{zhang2024attack},~\cite{huang2024robin},\cite{shukla2025waterflow}}, draw=gray!50, fill=citegray, text=black,text width=2cm]
      ]
      [{Key-based embedding\\(\cref{sec:iter_emb})}, draw=black, semithick, fill=key, text width=2.5cm, text=black
        [{\footnotesize~\cite{arabi2024hidden},~\cite{yang2024gaussian},~\cite{gunnPRC},~\cite{yang2025gaussian},~\cite{arabi2025seal}}, draw=gray!50, fill=citegray, text=black, text width=3cm]
      ]
    ]
  ]
\end{forest}}
\caption{Taxonomy of AI-generated image watermarking methods.}
\label{fig:taxonomy_ai_wm}
\Description{Taxonomy of AI-generated image watermarking methods.}
\end{figure}

\subsection{Fine-tuning-based Watermarking}\label{sec:fine-tune_wm}

\begin{figure}[t]
    \centering
    \includegraphics[width=0.7\linewidth]{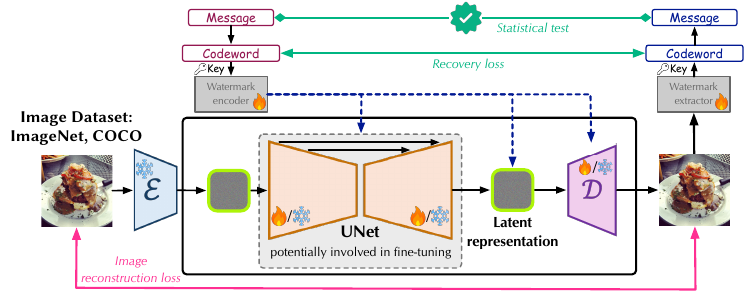}
    \caption{Training workflow of fine-tuning-based watermarking methods.}
    \label{fig:fine-tunedwatermark}
    \Description{A diagram showing the training pipeline of fine-tuning-based watermarking methods.}
\end{figure}

This category of in-generation watermarking includes methods that embed messages by directly modifying, retraining, or fine-tuning some components of the generative model. This pipeline, as defined in \Cref{eq:fine-tune_pipe}. Unlike post-hoc methods, it does not operate on a cover image ($I = \varnothing$), but instead starts from the text prompt $y$ and the initial noise $\mathbf{z}_T$ owing to the requirements of the generative model $\mathsf{Gen}$. The embedding locus $\Lambda$ is the model parameters $\theta$ themselves. In these methods, the message or its corresponding generated watermark signal becomes an intrinsic characteristic of the model’s output distribution. Specifically, these methods can modify the latent representations of LDM indirectly through parameter updates, resulting in watermarks that are inherently integrated into the generated outputs via the fine-tuned components. The overall workflow of these methods is illustrated in \Cref{fig:fine-tunedwatermark}.

\begin{equation}\label{eq:fine-tune_pipe}
\begin{aligned}
&m \xrightarrow{\ \mathsf{Encode}\ } w
    \xrightarrow{\ \mathsf{Modulate}_K\ } c_K(w) \xrightarrow{\mathsf{Embed}_\Lambda\in\{\mathsf{feature\;maps},\theta\}\big( I=\emptyset, y, \mathbf z_T, \theta; K, w, \rho \big)\ } I_w \xrightarrow{\ A\in\mathcal A\ } \\
    &I'\xrightarrow{\mathsf{Extract}(I'; K, \theta)} \hat m \xrightarrow{\ \mathsf{Verify}(\hat m, m)\ } \{\mathit{false},\mathit{true}\}.
\end{aligned}
\end{equation}

Fernandez \textit{et al.}~\cite{fernandez2023stable} proposed Stable Signature, the most representative AI-generated watermarking method that integrates signature embedding directly into the image generation process by fine-tuning model parameters without altering the LDM architecture. Initially, they pre-trained a watermark encoder $\mathcal{W}_\text{exc}$ and an extractor $\mathcal{W}_\text{ext}$ based on \Cref{eq:hidden_loss} to facilitate the embedding and extraction of keys. Secondly, they fine-tuned the VAE decoder $\mathcal{D}$ within the LDM, such that each generated image, when passed through the pre-trained watermark extractor $\mathcal{W}_{\text{ext}}$, yields a target binary key. The extractor’s output is used to compute a $\mathcal{L}_\text{BSE}$ loss function, which in turn guides the fine-tuning of $\mathcal{D}$. Therefore, the objective function of Stable Signature can be defined as:
    \begin{equation}
    \begin{aligned}
        &\mathcal{L}(\theta) =  \mathcal{L}_\text{BCE}(w,\mathcal{W}_\text{ext}(I_w)) + \alpha\mathcal{L}_\text{IMG}(I,I_w)), \quad\textit{where}\;I_w = \mathcal{D}(\mathcal{\mathbf{z}}_0),\;\theta=\theta_\mathcal{D}.
    \end{aligned}
    \end{equation}
    
Notably, Stable Signature exemplifies a model-targeted watermarking technique designed to trace the origin of generated content by identifying the specific generative model responsible for image synthesis and providing verifiable attribution. Its primary objective is to protect the IP and ownership rights associated with generative models.

In contrast to Stable Signature~\cite{fernandez2023stable}, Xiong \textit{et al.}~\cite{xiong2023flexible} proposed a secure and scalable watermarking architecture for LDMs that embeds user-specific identity information directly into the latent space. Their approach introduces a learnable user embedding module that injects messages into latent representations through linear interpolation, enabling a shared model architecture without requiring per-user fine-tuning. Within this framework, the VAE decoder $\mathcal{D}$ is subsequently fine-tuned to align with the embedded identities. Unlike model-targeted watermarking, which attributes content to a particular generative model, user-targeted watermarking~\cite{kim2024wouaf,min2024watermark} focuses on identifying the user of a GenAI model, who creates or distributes AI-generated images.

WOUAF~\cite{kim2024wouaf} is another user-target framework for LDM that facilitates user attribution while preserving image quality. It leverages a weight modulation technique applied to a pre-trained LDM. Each user’s unique message is first mapped to an intermediate representation, which modulates the weights of the VAE decoder via an affine transformation, thereby embedding the message directly into the generation process.

SleeperMark, proposed by Wang et al.~\cite{wang2024sleepermark}, fine-tunes the U-Net instead of the VAE decoder via a two-stage process. First, a watermark encoder and extractor are trained to encode messages into a fixed residual. Then, during U-Net fine-tuning, an optimization-based trigger mechanism embeds the residual into the output latent $\mathbf{z}_0$ when a prompt contains a specific trigger word. Prompts without the trigger yield outputs consistent with the original model.

\subsection{Initial Noise-based Watermarking}\label{sec:ini_wm}

Initial noise-based watermarking represents a significant advancement in in-generation watermarking techniques. Unlike fine-tuning-based watermarking, it embeds the message directly at the initial stage of the image generation process. The core idea leverages the intrinsic structure of diffusion models, which synthesize images by progressively denoising an initial noise sample, typically drawn from a standard Gaussian distribution denoted as $\mathbf{z}_{T}$ (as described in \Cref{sec:dm}). At the pipeline level, this class of methods sets $\Lambda=\mathbf{z}_T$, freezes all other parameters as in \Cref{eq:fine-tune_pipe}, and does not require any additional DNNs to implement the watermark encoder or extractor.

\begin{figure}[t]
    \centering
    \includegraphics[width=0.9\linewidth]{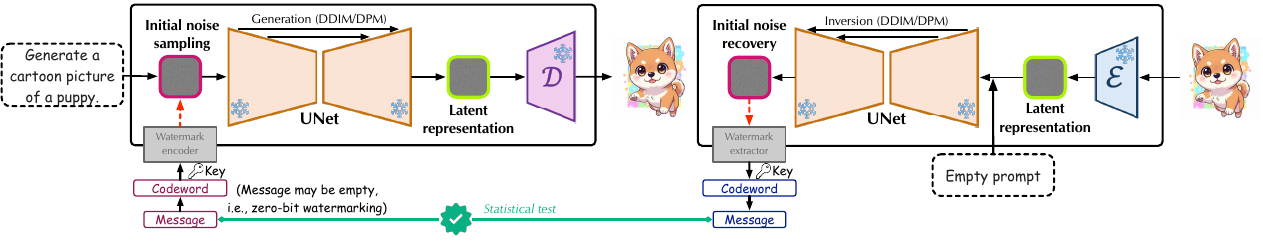}
    \caption{Initial noise-based watermarking method's typical workflow.}
    \label{fig:initialnoisewatermark}
    \Description{A diagram showing the typical workflow of an initial noise-based watermarking method. The process begins with a binary message that is encoded into a watermark signal. This signal is injected into the initial noise vector of a generative diffusion model. The model then generates a watermarked image that preserves the semantic content of the original prompt while embedding the watermark.}
\end{figure}

Initial noise-based watermarking subtly modifies the initial noise to embed the message or a distinctive watermark pattern, as shown in \Cref{fig:initialnoisewatermark}. The watermarked $\mathbf{z}_T$ is then propagated through the iterative denoising process, influencing the generation trajectory such that the watermark persists in the final output. For watermark verification, these methods generally perform an inverse diffusion process (e.g., DDIM~\cite{song2020denoising} or DPM-Solver$++$~\cite{lu2025dpm,hong2024exact}) to reconstruct an exact approximation of the watermarked initial noise from the generated image. The recovered noise is subsequently analyzed for the embedded pattern or compared against original messages to verify the watermark’s presence and authenticity. As shown in \Cref{fig:taxonomy_ai_wm}, depending on how the intial noise modification is introduced, existing approaches can be broadly categorized into direct embedding, iterative embedding, and key-based embedding frameworks, each offering different trade-offs in robustness, security, and design complexity.

\subsubsection{Direct Embedding} \label{sec:dir_emb}
Direct embedding, one of the initial-noise-based watermarking methods, injects watermark signals directly into the initial noise in a single step, without requiring iterative optimization or training. TreeRing~\cite{wen2023tree} is a pioneering method that embeds a watermark signal guided by concentric ring patterns (i.e., our defined $c_K(w)$) into the Fourier domain of the initial noise and can be described as:
\begin{equation}
\begin{aligned}
\mathcal{F}(\mathbf{z}_T^w)_i \sim 
\begin{cases}
c_K(w)_i & \text{if } w_i = 1 \\
\mathcal{N}(0,1) & \text{if } w_i = 0
\end{cases}
\end{aligned}
\end{equation}
where $w$ serves as a binary circular mask that determines the watermark embedding locations. Although TreeRing is a zero-bit watermarking scheme without an explicitly defined message or codeword, the structured nature of $w$ allows it to be interpreted as an approximate codeword, guiding the construction of the watermark signal. $c_K(w) \in \mathbb{C}^{|w|}$ is sampled in a complex space. Then, $\mathsf{Gen}$ uses $\mathbf{z}^w_T$ to generate the watermarked image $I_w$. The TreeRing's watermark extraction and verification relies on reconstructing the initial latent using DDIM inversion, and the detection score $S$ in $\mathsf{Verify}$ can be computed as follows:
\begin{equation}
\begin{aligned}
    &\hat{\mathbf{z}^w_T} = \mathsf{DDIM}_{0\rightarrow T}(\mathcal{E}(I_w)),\quad S =  \sum_{i=1} w_i \cdot \left| c_K(w)_i - \mathcal{F}(\hat{\mathbf{z}}_T^w)_i \right|.
\end{aligned}
\end{equation}

TreeRing~\cite{wen2023tree} is a training-free watermarking method that demonstrates superior performance compared to fine-tuning-based approaches. However, TreeRing is a zero-bit watermarking scheme, meaning it can only indicate the presence of a watermark. Moreover, studies have shown that TreeRing introduces deviations from the standard Gaussian distribution, which may degrade image quality and reduce generation diversity. To address these limitations, Ci \textit{et al.}~\cite{ci2024ringid} proposed RingID, a multi-key watermarking technique that employs multi-channel heterogeneous watermarks, allowing multiple TreeRing-like watermark types to be embedded into distinct channels of the initial noise. Their implementation integrates an enhanced TreeRing pattern with Gaussian noise, resulting in a watermark whose distribution aligns with that of the original noise. This design provides theoretically unlimited capacity and strong resilience against non-geometric attacks.

\begin{table}[t]
\centering
\caption{Summary of AI-generated image watermarking methods.}

\rowcolors{2}{gray!10}{white}

\resizebox{\textwidth}{!}{%
\begin{tabular}{>{
\centering\arraybackslash}m{1.2cm}@{} 
>{\centering\arraybackslash}m{1.2cm}
>{\centering\arraybackslash}m{1.2cm}
>{\centering\arraybackslash}m{2.5cm}
>{\centering\arraybackslash}m{2cm}@{} 
>{\centering\arraybackslash}m{1.5cm}@{} 
>{\centering\arraybackslash}m{2.5cm} 
>{\centering\arraybackslash}m{6.5cm}
}
\toprule
\multicolumn{1}{l}{\textbf{Method}} &\textbf{Sec.} &\textbf{Rob.} & \makecell{\textbf{Modified comp.}\\\textbf{ of the SDM}} & \makecell{\textbf{Watermark}\\ \textbf{format}} & \makecell{\textbf{Target}} & \makecell{\textbf{SDM} \\ \textbf{Ver.}}  & \textbf{Used Dataset} \\
\midrule

  \rowcolor{blue!10}\multicolumn{8}{c}{\textbf{Fine-tuning-based Watermarking}} 
\\ \midrule
\cite{fernandez2023stable} &\whitecircle &\halfblackcircle   &$\mathcal{D}$&B.&M. &SD-2.1 &COCO, ImageNet  \\

\cite{xiong2023flexible} &\halfblackcircle &\halfblackcircle &$\mathcal{D}$&B.&U. &$-$ &LAION-400B  \\
\cite{kim2024wouaf} &\whitecircle &\halfblackcircle  &$\mathcal{D}$&B.&U. &SD-2 &COCO, LAION-400B \\

\cite{wang2024sleepermark} &\whitecircle &\halfblackcircle &$\boldsymbol{\epsilon}_\theta$&B.&M. &SD-1.4 &COCO, SD-Prompts  \\

\cite{feng2024aqualora} &\whitecircle &\halfblackcircle &$\boldsymbol{\epsilon}_\theta$ &B.&M. &SD-1.5 & COCO, SD-Prompts \\
\cite{ciwmadapter} &\whitecircle &\halfblackcircle &$\mathcal{D}$ &B.&U. &SD-2.1 &COCO  \\
\cite{lu2024robust} &\whitecircle &\halfblackcircle &$\theta_{\mathsf{Gen}}$ &B. &U.& SDXL-Turbo &OpenImages, UltraEdit, Flickr, COCO \\

 \midrule
\rowcolor{red!10} \multicolumn{8}{c}{\textbf{Initial Noise-based Watermarking}}\\
 \midrule
\cite{wen2023tree} &\whitecircle &\halfblackcircle  &$-$ &Z. &U.  &SD-2  &COCO, ImageNet\\
\cite{zhu2024flexible} &\whitecircle &\halfblackcircle &$-$ &B. &M. &SD-2 &COCO, SD-Prompts \\
\cite{ci2024ringid} &\whitecircle &\halfblackcircle  &$-$ &B. &U. &SD-2.1 &COCO \\

\cite{zhang2024attack}  &\whitecircle &\halfblackcircle &$-$ &Z. &U.  &SD-2  & COCO, DiffusionDB, WikiArt \\

\cite{huang2024robin} &\whitecircle &\halfblackcircle &$-$ &Z. &U. &SD-2  &ImageNet, SD-Prompts \\ 
\cite{shukla2025waterflow} &\whitecircle &\halfblackcircle &$-$ &Z. &U. &$-$ &COCO, DiffusionDB, WikiArt  \\

\cite{arabi2024hidden} &\halfblackcircle &\halfblackcircle &$-$ &Z. &U. &SD-2.1, SD-1.4 &COCO, SD-Prompts \\
\cite{yang2024gaussian} &\halfblackcircle &\halfblackcircle & $-$ &B. &U.  & SD-1.4, 2, 2.1 &SD-Prompts \\
\cite{gunnPRC} &\halfblackcircle &\halfblackcircle &$-$ &B. &U. &SD-2.1  &COCO, SD-Prompts \\
\cite{arabi2025seal}  &\halfblackcircle &\halfblackcircle  &$-$ &B. &U. &SD-2.1-base  &COCO, SD-Prompts \\
\cite{yang2025gaussian} &\blackcircle &\blackcircle & $-$ &B. &U. & SD-2.1  &SD-Prompts
\\
\bottomrule
\end{tabular}
}
\begin{tablenotes}
\footnotesize
\item \textit{Sec.} and \textit{Rob.} indicate whether the watermarking method is secure and robust, respectively. \text{\blackcircle} indicates full support, \text{\whitecircle} means no support, and \text{\halfblackcircle} denotes partial support. \textit{Modified comp. of the SDM} refers to components that have been altered or fine-tuned within the SDM. \textit{B.} or \textit{Z.} denote the message format, i.e., binary string or zero-bit. \textit{M. U.} indicates whether the method is model-targeted or user-targeted. \textit{SDM Ver.} specifies the version of the SDM used.
\end{tablenotes}

\label{tab:existingAIwatermark}

\end{table}

\subsubsection{Iterative Embedding} \label{sec:iter_emb}
Iterative embedding methods inject the watermark into the initial noise through multi-step optimization to improve the image quality compared to direct embedding methods, which lack explicit control over image quality. Zhang \textit{et al.}~\cite{zhang2024attack} introduced ZoDiac, which embeds a watermark into an existing image by optimizing the initial noise of a pre-trained and frozen LDM. Specifically, it first derives the noise corresponding to the target image via DDIM inversion, embeds a TreeRing-like concentric watermark in $\mathbf{z}_T$'s Fourier domain, and iteratively updates it by minimizing the reconstruction loss between the generated and original images. 

ROBIN~\cite{huang2024robin} adopts an iterative optimization strategy to embed watermarks during intermediate steps of the DDIM sampling process, rather than altering the initial noise. It implants the watermark into the frequency domain at a selected step and introduces a hiding-prompt-guidance signal to steer the remaining denoising, enhancing imperceptibility. Watermarks and guidance signals are co-learned via adversarial optimization to balance robustness and fidelity.

A growing number of works have concentrated on initial-noise-based watermarking methods. Recent studies such as~\cite{min2024watermark,zhu2024flexible,lei2024diffusetrace,arabi2025seal,shukla2025waterflow} exemplify this research direction. Although these approaches differ in their specific design strategies, they share a common objective: To construct watermarked initial noise that remains statistically consistent with the original Gaussian distribution used in LDMs. This principle is crucial for preserving the imperceptibility of the embedded watermark and the fidelity of the generated images.

\subsubsection{Key-based Embedding}

In the aforementioned AI-generated image watermarking methods, no cryptographic key $K$ is used, i.e., $K = \emptyset$. Latest works explicitly incorporate cryptographic primitives into the watermarking pipeline. Concretely, the message $m$ is first encoded into a codeword $w$, which is then embedded into the initial noise using a secret key $K$. The resulting watermarked initial noise $\mathbf{z}_T$ is exactly our defined $c_K(w)$. By integrating cryptographic operations into the modulation and embedding stages, these approaches aim to strengthen the overall security of the watermarking system.

Yang et al.~\cite{yang2024gaussian} proposed Gaussian Shading, a distribution-preserving sampling method that embeds watermarks into initial noise. The message $m$ is encoded and encrypted with a stream cipher key $K$, then used to sample values from equally probable micro-intervals of the Gaussian space. This ensures the noise encodes the watermark without altering its distribution. Gaussian Shading++~\cite{yang2025gaussian} extends this by enabling a fixed key to generate pseudorandom bit streams across generations, enhancing scalability and key management.

Gunn \textit{et al.}~\cite{gunnPRC} employed a cryptographic primitive known as the Pseudorandom Error-correcting Code (PRC), which is also applied at the stage of initial noise sampling. In this approach, the pseudorandom $c_K(w)$ consisting of $+1$ and $-1$ values is generated using a cryptographic key $K$ and the message $m$, and this keyed carrier determines the sign of each element in a sampled standard Gaussian noise, thereby producing the watermarked initial noise $\mathbf{z}^w_T$. Because the generated $c_K(w)$ are computationally indistinguishable from true randomness, the resulting watermarked initial noise remains statistically identical to standard Gaussian noise. This theoretical indistinguishability guarantees that the embedded message introduces no perceptible or computationally detectable degradation in image quality.

\begin{figure}[t]
    \centering
    \includegraphics[width=0.5\linewidth]{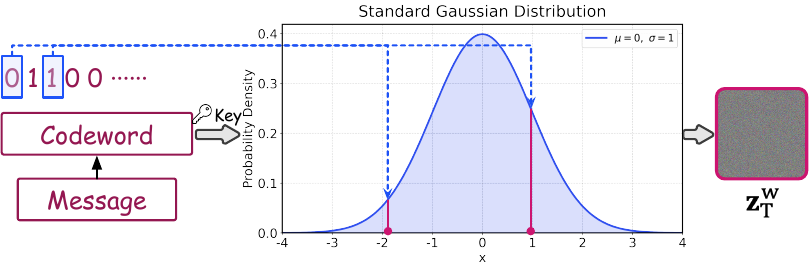}
    \caption{Key-driven sampling of standard Gaussian noise from a codeword.}
    \label{fig:sampling}
    \Description{A diagram illustrating the process of generating a standard Gaussian noise value from a binary codeword using a secret key.}
\end{figure}

The core idea behind \cite{yang2024gaussian} and \cite{gunnPRC} is to start from a message $m$ and, under the control of a secret key $K$, sample a standard Gaussian distribution, as illustrated in Fig.~\ref{fig:sampling}. Conceptually, both methods are built around a key-dependent, approximately reversible mapping from bits $(0,1)$ to a sample point of the standard Gaussian. In the $\mathsf{Embed}$, codeword $w$ is constructed from $m$, which is then transformed under the key $K$ into a set of Gaussian samples that define the watermarked initial noise $\mathbf{z}^w_T$. During verification, the same key $K$ enables the inverse mapping from the estimated initial noise back to the underlying bit sequence, from which the original message $m$ can be recovered.

Compared to other initial-noise-based methods, PRC watermarking offers higher capacity. It encodes information jointly across the entire $c \times h \times w$ latent space. Zero-bit methods such as TreeRing\cite{wen2023tree} and RingID\cite{ci2024ringid} scale capacity with the number of concentric rings. Gaussian Shading~\cite{yang2024gaussian,yang2025gaussian} uses per-dimension interval partitioning. In contrast, PRC embeds the secret key as a full-length vector, achieving up to 2500 bits while preserving imperceptibility and statistical consistency.

In summary, we reviewed the existing works in AI-generated image watermarking, including fine-tuning-based methods and initial noise-based methods. Fine-tuning-based methods embed watermarks by directly modifying or retraining key components of the generative model (see~\Cref{tab:existingAIwatermark}), making the watermark an intrinsic property of the model. This paradigm can be designed for specific objectives such as model copyright protection (e.g., Stable Signature~\cite{fernandez2023stable}) or user identity tracing (e.g., WOUAF~\cite{kim2024wouaf}). However, this approach still faces several limitations: (i) it entails high computational complexity, as model fine-tuning requires considerable computing resources; (ii) its scalability is limited, making it difficult to generalize across model architectures; and (iii) it may leave residual watermark artifacts in the pixel space, reducing imperceptibility and potentially revealing the presence of watermarking. In contrast, initial-noise-based watermarking methods present a more flexible and adaptable alternative. As shown in \Cref{tab:existingAIwatermark}, a range of techniques, from early approaches like TreeRing~\cite{wen2023tree} to high-capacity methods like PRC~\cite{gunnPRC}, embed watermarks by modifying the initial noise of LDM, allowing the watermark signal to persist throughout the denoising process. This ``training-free'' property greatly simplifies deployment while enabling exceptionally high watermark capacity. However, even the latest key-based initial-noise watermarking methods still suffer from limited robustness \cite{francati2025coding}.

%Fig.~\ref{fig:main_gallery} compares visual results from various generative watermarking methods, highlighting differences in quality and prompt alignment. The key challenge remains embedding secure and robust watermark signals without compromising image fidelity or diversity, a progressive area of ongoing research.

\section{Evaluation of AI-generated Image Watermarking}\label{sec:evalmetric}

The effectiveness of AI-generated image watermarking systems should be assessed using standardized metrics that capture multiple dimensions of performance. This section outlines the key evaluation criteria, focusing on four primary aspects: (1) the visual quality of watermarked images, (2) the information-carrying capacity of the embedded watermark, (3) the statistical reliability of watermark detection and user attribution, and (4) existing benchmarks within this field.

\subsection{Visual Quality}
To comprehensively evaluate the quality, fidelity, and semantic consistency of AI-generated images, a variety of quantitative metrics are commonly employed, including Structural Similarity Index (SSIM)~\cite{wang2004image}, Peak Signal-to-Noise Ratio (PSNR)~\cite{setiadi2021psnr}, Fréchet Inception Distance (FID)~\cite{heusel2017gans}, LPIPS~\cite{zhang2018unreasonable_LPIPS}, and the Contrastive Language–Image Pretraining (CLIP) score~\cite{radford2021learning}. 

SSIM and PSNR assess image similarity by measuring structural integrity and pixel-level distortion, respectively, but may not reflect perceptual or semantic differences. To overcome this, FID quantifies distributional similarity in feature space, and LPIPS evaluates perceptual differences using deep network embeddings, offering better alignment with human judgment. CLIP score measures semantic alignment between images and text prompt, making it especially relevant for text-to-image tasks. Collectively, these metrics form a comprehensive framework for evaluating visual fidelity and semantic consistency in generative models.

Nevertheless, objective metrics alone are insufficient to assess watermark imperceptibility, as they fail to capture the nuances of the Human Visual System (HVS). For instance, methods with high LPIPS scores may still introduce visible artifacts in smooth regions. Thus, subjective visual assessment remains a reliable approach. Providing qualitative examples has become standard practice~\cite{fernandez2022watermarking,xiong2023flexible,kim2024wouaf,wang2024sleepermark,wen2023tree,ci2024ringid,zhang2024attack,yang2024gaussian}, serving as a crucial complement to quantitative evaluation.

\subsection{Capacity}

Watermark capacity refers to the amount of information that can be embedded within a digital file, typically measured in bits~\cite{wan2022comprehensive,ren2024sok,luo2025digital}. Some approaches, such as TreeRing~\cite{wen2023tree}, employ zero-bit watermarks, which merely indicate the presence or absence of a watermark. In contrast, other methods~\cite{fernandez2023stable,kim2024wouaf,xiong2023flexible,yang2024gaussian,yang2025gaussian,gunnPRC} utilize multi-bit binary strings to encode richer information. In image watermarking, capacity is often influenced by the dimensionality of the latent space and additional factors such as the channel diffusion factor, spatial diffusion factor, and code rate~\cite{yang2024gaussian}. Crucially, capacity is not an isolated metric. It is intrinsically linked with other key properties of watermarking systems, including imperceptibility, robustness, and computational cost. Increasing capacity often introduces trade-offs: it may degrade visual or semantic quality, elevate the bit error rate, or weaken the resilience of the watermark against distortions and attacks. Therefore, the development of effective watermarking techniques hinges on the ability to balance the inherent trade-offs between security, robustness, visual quality, and capacity~\cite{zhao2024sok}.

\begin{table}[t]
\centering
\caption{The quality evaluation metrics for different AI-generated watermarking methods.~\textcolor{green}{$\blacktriangle$} indicates that higher values of the metric are preferable, while~\textcolor{red}{$\blacktriangledown$} indicates that lower values are preferable.}
%\rowcolors{2}{gray!10}{white}

\resizebox{\textwidth}{!}{%

\begin{tabular}{>{\raggedright\arraybackslash}m{4cm}@{} 
>{\centering\arraybackslash}m{4cm}
>{\raggedright\arraybackslash}p{11cm}}
\toprule

\textbf{Evaluation Dimension} & \textbf{Metric}   & \textbf{Method} \\
\midrule

\multirow{6}{*}{\makecell[{{p{3.5cm}}}]{\textbf{Visual Quality}}} 
&   PSNR\textcolor{green}{$\blacktriangle$}  &\cite{fernandez2023stable},~\cite{xiong2023flexible},~\cite{zhang2024attack},~\cite{arabi2024hidden},~\cite{huang2024robin},~\cite{ciwmadapter},~\cite{lu2024robust},~\cite{shukla2025waterflow}  \\
&   SSIM\textcolor{green}{$\blacktriangle$} &\cite{fernandez2023stable},~\cite{xiong2023flexible},~\cite{zhang2024attack},~\cite{arabi2024hidden},~\cite{huang2024robin},~\cite{ciwmadapter},~\cite{lu2024robust},~\cite{shukla2025waterflow}   \\
&   FID\textcolor{red}{$\blacktriangledown$} &\cite{fernandez2023stable},~\cite{xiong2023flexible},~\cite{kim2024wouaf},~\cite{wang2024sleepermark},~\cite{wen2023tree},~\cite{yang2024gaussian},~\cite{yang2025gaussian},~\cite{gunnPRC},~\cite{arabi2024hidden},~\cite{huang2024robin},~\cite{feng2024aqualora},~\cite{ciwmadapter},~\cite{lu2024robust},~\cite{zhu2024flexible}     \\
&   LPIPS\textcolor{red}{$\blacktriangledown$}&\cite{zhang2024attack},~\cite{shukla2025waterflow},~\cite{gunnPRC},~\cite{lu2024robust}     \\
&   CLIP score\textcolor{green}{$\blacktriangle$}&\cite{kim2024wouaf},~\cite{wen2023tree},~\cite{ci2024ringid},~\cite{yang2024gaussian},~\cite{yang2024gaussian},~\cite{wang2024sleepermark},~\cite{gunnPRC},~\cite{arabi2025seal},~\cite{huang2024robin},~\cite{zhu2024flexible}       \\

&   Visualization &\cite{fernandez2023stable},~\cite{xiong2023flexible},~\cite{wang2024sleepermark},~\cite{wen2023tree},~\cite{ci2024ringid},~\cite{zhang2024attack},~\cite{yang2024gaussian},~\cite{yang2024gaussian},~\cite{gunnPRC}      \\

\midrule
\multirow{3}{*}{\textbf{Capacity}}
&  \multirow{1.5}{*}{Multi-bit watermarking\textcolor{green}{$\blacktriangle$}}  &\texttt{48-bit}~\cite{fernandez2023stable,wang2024sleepermark,ciwmadapter,feng2024aqualora,zhu2024flexible};\,
  \texttt{100-bit}~\cite{lu2024robust};\,
  \texttt{128-bit}~\cite{xiong2023flexible,kim2024wouaf};\,
  \texttt{256-bit}~\cite{yang2024gaussian,yang2025gaussian};\,
  \texttt{2500-bit}~\cite{gunnPRC}  \\

& Zero-bit watermarking  &\cite{wen2023tree},~\cite{ci2024ringid},~\cite{huang2024robin},~\cite{zhang2024attack},~\cite{xian2024raw},~\cite{arabi2024hidden},~\cite{shukla2025waterflow} \\

\midrule
\multirow{2}{*}{\textbf{Detection and Attribution}} 
&   BA\textcolor{green}{$\blacktriangle$}  &\cite{fernandez2023stable},~\cite{xiong2023flexible},~\cite{kim2024wouaf},~\cite{wang2024sleepermark},~\cite{yang2024gaussian},~\cite{yang2025gaussian},~\cite{feng2024aqualora},~\cite{ciwmadapter},~\cite{lu2024robust},~\cite{zhu2024flexible}  \\
& $\mathrm{TPR@\varepsilon FPR}$\textcolor{green}{$\blacktriangle$}    &\cite{fernandez2023stable},~\cite{wang2024sleepermark},~\cite{wen2023tree},~\cite{ci2024ringid},~\cite{zhang2024attack},~\cite{yang2024gaussian},~\cite{yang2025gaussian},~\cite{gunnPRC},~\cite{arabi2024hidden},~\cite{feng2024aqualora},~\cite{ciwmadapter},~\cite{lu2024robust}~\cite{zhu2024flexible},\cite{huang2024robin},~\cite{arabi2025seal}   \\

\bottomrule
\end{tabular}}

\label{tab:quality_metrics}
\end{table}

\subsection{Detectability} \label{sec:dete_metric}
A complete watermarking system typically comprises two fundamental components: watermark detection and watermark attribution with multi-user traceability. The detection module determines whether a given image contains an embedded watermark, serving as the primary mechanism for distinguishing AI-generated watermarked images from unwatermarked ones. In contrast, the attribution module extends this capability by decoding user-specific identifiers or ownership information embedded within the watermark, thereby enabling provenance tracking and accountability in multi-user settings. Together, these two functions ensure that the system achieves reliable detection with a low false-positive rate while supporting fine-grained source attribution, which is a prerequisite for responsible and transparent governance of generative AI technologies.

\subsubsection{Watermark Detection}
The objective of watermark detection is to accurately determine whether an image contains an embedded message while minimizing false alarms. A reliable $\mathsf{Verify}$ must ensure that ``vanilla images", that is, naturally generated or unwatermarked images, are not erroneously classified as watermarked, thereby avoiding false positives~\cite{francati2025coding}. The $\mathrm{FPR}$ serves as a key metric for assessing watermark detectability. It measures the likelihood that a non-watermarked image is incorrectly identified as containing a watermark and is formally defined as:

\begin{myDef}
$\mathrm{FPR}$ measures the probability that a non-watermarked image is incorrectly detected as containing a watermark.
\begin{equation}
    \Pr \big[\mathsf{Verify}(\hat m, m) \to \mathit{true} \big]\leq \varepsilon.
\end{equation}
\end{myDef}

The verification function $\mathsf{Verify}$ is typically based on a decision threshold $\tau$, which is determined by the detection score between the extracted message $\hat{m}$ and the original $m$. Following the statistical testing framework in~\cite{fernandez2023stable}, a standard procedure is to fix an acceptable false-positive rate $\varepsilon$ (e.g., $10^{-4}$~\cite{guo2024ai,jiang2023evading}) and estimate the score distribution under the null hypothesis $H_0$ (i.e., the image does not contain a watermark). The threshold $\tau$ is then selected such that
\begin{equation}\label{eq:fpr}
\mathrm{FPR}(\tau)=\Pr(S \geq \tau \mid H_0) = \varepsilon,
\end{equation}
where $S$ denotes the detection score, which is commonly measured as Bit Accuracy (BA) in some prior works \cite{fernandez2023stable,wu2024robustness,jiang2023evading,kassis2025unmarker}. It measures the amount or the proportion of correctly decoded bits in the extracted message $\hat{m}$ relative to the ground-truth message $m$. In practice, the threshold can be derived from the cumulative distribution function of a binomial model~\cite{fernandez2023stable}.  
\begin{equation}
\mathrm{BA}(\hat{m}, m) 
= \frac{1}{|m|}\sum_{i=1}^{|m|} \mathbf 1[\hat{m}_i = m_i].
\end{equation}

To assess detection reliability, the metric $\mathrm{TPR@\varepsilon FPR}$, true positive rate at a fixed FPR, is widely used. This measure quantifies the proportion of watermarked samples correctly detected while ensuring that the $\mathrm{FPR}$ remains below a specified threshold $\varepsilon$. Compared to metrics such as BA, $\mathrm{TPR@\varepsilon FPR}$ provides a more realistic indicator of the detector's operational robustness, since a high BA may still correspond to an unacceptably high number of false alarms under stringent $\mathrm{FPR}$ constraints.

\subsubsection{Watermark Attribution and Traceability}
 Watermarking systems often support \emph{attribution} or \emph{multi-user traceability}, enabling the identification of the specific user or model instance responsible for generating the content. In a multi-user scenario, each message encodes a unique identifier, and verification must be performed against $N$ distinct hypotheses, each corresponding to $N$ registered users.  

Based on~\Cref{eq:fpr}, the global $\mathrm{FPR}$ across all users can be presented as:
\begin{equation}
\mathrm{FPR}(\tau, N)=1-(1-\mathrm{FPR}(\tau))^N.
\end{equation}

This equation illustrates how the probability of false attribution increases as the number of users grows. For example,~\cite{fernandez2023stable} sets $\tau=44$ for $|m|=48$, $\mathrm{FPR}=10^{-6}$, and $N=1000$, assuming a Bernoulli model for bit-level independence.

\subsection{Benchmarks and Tools}
To support the standardized implementation and evaluation of watermarking techniques, several benchmarks and toolkits have recently been introduced, among which WAVES~\cite{an2024waves} and MarkDiffusion~\cite{pan2025markdiffusion} are particularly notable.

WAVES~\cite{an2024waves} represents the first comprehensive benchmark designed to systematically evaluate the robustness of image watermarking methods. It establishes a standardized evaluation workflow that addresses the inconsistencies present in prior studies. The framework comprises a diverse and extensive suite of stress tests, encompassing not only conventional image distortions but also regeneration-based and adversarial attacks. By extensively testing three representative watermarking methods, namely, Stable Signature~\cite{fernandez2023stable}, TreeRing~\cite{wen2023tree}, and StegaStamp~\cite{tancik2020stegastamp}, WAVES exposes previously unrecognized vulnerabilities in each approach. Moreover, it jointly assesses two critical dimensions: (1) the effectiveness of watermark detection after attacks and (2) the degree of image quality degradation caused by these attacks. By integrating multiple quality metrics and introducing innovative ``Performance vs. Quality” 2D plots, WAVES delivers a unified and realistic analysis of robustness trade-offs and hidden weaknesses in existing watermarking systems.

Complementing such benchmarks, MarkDiffusion~\cite{pan2025markdiffusion} provides an open-source toolkit for AI-generated image watermarking in LDMs. It offers a unified implementation framework that supports numerous initial-noise-based watermarking methods~\cite{wen2023tree,ci2024ringid,huang2024robin,yang2024gaussian,gunnPRC}, streamlining experimentation through user-friendly interfaces. In addition, MarkDiffusion includes a comprehensive evaluation suite that measures watermark detectability, visual quality, and robustness, alongside an interactive visualization module designed to help both researchers and the general public better understand the underlying mechanisms of AI-generated image watermarking.

\subsection{Evaluation Metrics in Existing Watermarking Methods}
% In this subsection, we have detailed the essential metrics for evaluating watermarking methods applied to AI-generated images, and introduce a summary in {tab:quality_metrics}. We believe that a comprehensive empirical quality validation must first consider visual quality, using a combination of quantitative metrics, such as SSIM, PSNR, FID, and LPIPS, and subjective visual evaluations to ensure the watermark's imperceptibility. Secondly, watermark capacity, measured in bits, defines the amount of information that can be embedded, ranging from zero-bit presence detection to multi-bit strings for user IDs. Finally, the reliability of watermark detection is essential, fundamentally assessed using the $\mathrm{FPR}$ to control for incorrect detections on unwatermarked content. The detection threshold is set to meet a strict, predefined $\mathrm{FPR}$, with metrics like $\mathrm{TPR@FPR}$ and BA used to measure performance under these constraints.

We summarize key evaluation metrics for AI-generated image watermarking in \Cref{tab:quality_metrics}, covering three main aspects: visual quality, watermark capacity, and detection reliability. For visual quality, the CLIP score is commonly used by initial-noise-based methods~\cite{wen2023tree,ci2024ringid,yang2024gaussian,yang2025gaussian} to assess semantic alignment with prompts, as noise modifications may affect generation trajectories. FID is broadly adopted across both fine-tuning and noise-based approaches to measure distributional similarity to real images. Most works~\cite{fernandez2023stable,xiong2023flexible,wang2024sleepermark,wen2023tree,ci2024ringid,zhang2024attack,yang2024gaussian,gunnPRC} also include qualitative visualizations to complement objective metrics. Watermark capacity is evaluated via maximum message length, while BA and FPR are standard for assessing detection and attribution reliability.

\section{Attacks against AI-generated Image Watermarking}\label{sec:attack}

During online dissemination, AI-generated images are not only subject to degradation from common image processing operations but also face the threats of adversarial attacks launched by potential adversaries. As shown in \Cref{fig:taxonomy_attack}, these threats fall into two main categories based on \textit{attacker's objectives}: watermark removal attacks and watermark forgery attacks. In this section, we examine these security challenges in depth and provide a systematic analysis of the underlying principles and mechanisms behind these attacks. 

\definecolor{removalmain}{RGB}{191,148,228}
\definecolor{removalsub}{RGB}{223,199,247}

\definecolor{forgerymain}{RGB}{0,191,255}
\definecolor{forgerysub}{RGB}{180,235,255}

\definecolor{advgold}{RGB}{218,145,0}
\definecolor{advlight}{RGB}{245,210,130}

\definecolor{regenred}{RGB}{240,128,128}
\definecolor{regenlight}{RGB}{255,200,200}

\definecolor{citegray}{gray}{0.95}

\begin{figure}[t]
\centering
\tikzset{
  my node/.style={
    draw=black, semithick, align=center, text width=1.2cm, rounded corners=3,
  },
  my leaf/.style={
    draw=black, semithick, align=left, text width=8.5cm, rounded corners=3,
  }
}
\scalebox{0.85}{%
\forestset{
  every leaf node/.style={if n children=0{#1}{}},
  every tree node/.style={if n children=0{minimum width=1em}{#1}},
}
\begin{forest}
  nonleaf/.style={font=\scriptsize},
  for tree={
    every leaf node={my leaf, font=\tiny},
    every tree node={my node, font=\tiny, l sep-=4.5pt, l-=1pt},
    anchor=west, inner sep=2pt,
    l sep=10pt, s sep=3pt, fit=tight,
    grow'=east, edge={semithick},
    parent anchor=east, child anchor=west,
    if n children=0{}{nonleaf},
    edge path={\noexpand\path[draw,\forestoption{edge}]
      (!u.parent anchor) -- +(5pt,0) |- (.child anchor)\forestoption{edge label};},
    if={isodd(n_children())}{
      for children={
        if={equal(n,(n_children("!u")+1)/2)}{calign with current}{}
      }
    }{}
  }
  [{\textbf{Watermark Attack Methods}}, draw=black, semithick, fill=gray!15, text width=2cm, text=black
    [{\textbf{Removal Attacks\\(\cref{sec:removal_att})}}, draw=black, semithick, fill=removalmain!25, text width=2.5cm, text=black
      [{Regeneration Attacks\\(\cref{sec:rengen_att})}, draw=black, semithick, fill=regenlight, text width=2.5cm, text=black
        [{\footnotesize~\cite{zhao2024invisible},~\cite{an2024waves},~\cite{saberirobustness},~\cite{liu2024image}}, draw=gray!50, fill=citegray, text=black, text width=2.7cm]
      ]
      [{Adversarial Attacks\\(\cref{sec:adv_att})}, draw=black, semithick, fill=advlight, text width=2.5cm, text=black
        [{\footnotesize~\cite{jiang2023evading},~\cite{hu2024transfer},~\cite{hu2024stable},~\cite{an2024waves},~\cite{yang2024can},~\cite{kassis2025unmarker},~\cite{muller2025black}}, draw=gray!50, fill=citegray, text=black, text width=4.5cm]
      ]
    ]
    [{\textbf{Forgery Attacks\\(\cref{sec:forgery_att})}}, draw=black, semithick, fill=forgerymain!25, text width=2.5cm, text=black
      [{\footnotesize~\cite{yang2024can},~\cite{zhu2025optimization},~\cite{muller2025black},~\cite{jain2025forging},~\cite{dong2025imperceptible},~\cite{luo2025watermark}}, draw=gray!50, fill=citegray, text=black, text width=4cm]
    ]
  ]
\end{forest}}
\caption{Taxonomy of watermark attack methods.}
\label{fig:taxonomy_attack}
\Description{Taxonomy of watermark attack methods.}
\end{figure}

\subsection{Adversary Prior Knowledge and Capability}
Understanding the potential capabilities of attackers is essential for designing resilient watermarking systems. To systematically analyze these challenges, we follow the framework proposed in~\cite{zhao2024sok} and classify adversarial capabilities into two primary categories, based on the adversary’s level of interaction with the watermarking system and the resources available to them. In addition, we provide a consolidated comparison in \Cref{tab:attack_summary}, detailing each attack’s source, adversary capabilities, and targeted watermarking approaches. 

%We provide a foundation for structured security analysis in AI-generated image watermarking.

\begin{itemize}
    \item \textbf{Access to Data} \\
    This dimension reflects the adversary’s ability to access data that serves as the basis for analysis, comparison, and reverse engineering. Two primary cases can be distinguished:
    \begin{itemize}
        \item[\ding{192}] \textit{Watermarked Images:} Direct access to images generated by the target model with an embedded watermark. 
        \item[\ding{193}] \textit{Non-Watermarked Images:} Access to clean images generated by the same model without watermarking, which can be exploited for differential analysis.
    \end{itemize}

    \item \textbf{Access to System Internals and Oracles} \\
    This dimension reflects the degree of interaction with the underlying system components, representing progressively stronger adversarial capabilities:
    \begin{itemize}
        \item \textbf{\textit{Generative Model Access:}}
        \begin{itemize}
            \item[\ding{194}] \textit{Surrogate Model Access:} Utilization of a functionally similar (often lower-performing) model to regenerate content in order to bypass watermarking when the original model is inaccessible. 
            \item[\ding{195}] \textit{White-Box Access:} Full visibility and modification rights to the parameters and architecture of the generative model, representing the strongest level of model access.
        \end{itemize}
        \item \textbf{\textit{Watermark Embedding and Verification Mechanism Access:}}
        \begin{itemize}
            \item[\ding{196}] \textit{Embedding Mechanism Access}: The ability to infer the watermarking scheme type, or to understand the embedding process, without having direct access to the system internals.
            \item[\ding{197}] \textit{Verifier Oracle Access:} The ability to query modified images against the verification model and obtain binary feedback, facilitating iterative trial-and-error attacks.  
            \item[\ding{198}] \textit{Verifier Feedback Access:} Enhanced oracle capabilities that return more informative outputs (e.g., probability scores or extracted messages), which substantially improve the efficiency of attack optimization.  
            \item[\ding{199}] \textit{Chosen-Key Generation Access:} The most powerful embedding-level privilege, granting the ability to generate watermarked images under arbitrary messages and keys, thereby directly exposing the embedding logic.
        \end{itemize}
    \end{itemize}
\end{itemize}
\begin{table*}[!t]
\centering
\caption{Summary of Watermark Attack Methods.}
\label{tab:attack_summary}
\rowcolors{2}{gray!10}{white}
\resizebox{\textwidth}{!}{%
\begin{tabular}{
    >{\RaggedRight}m{4.5cm} % Attack Method Name
    >{\centering\arraybackslash}m{1cm} % Source Paper
    >{\centering\arraybackslash}m{2.5cm} % Capabilities
    >{\RaggedRight}m{8.53cm} % Targeted Watermarking Schemes
}
\toprule
\textbf{Attack Method } & 
\textbf{Source}& 
\textbf{Capabilities} & 
\textbf{Targeted Watermarking Method} \\
  \midrule
\rowcolor{blue!10}\multicolumn{4}{c}{\textbf{Removal attack}} 
\\ \midrule
WEvade-W-II & \cite{jiang2023evading} & \ding{198} & \cite{zhu2018hidden}, \cite{zhang2020udh} \\

WEvade-B-Q & \cite{jiang2023evading} &\ding{192}, \ding{197} & \cite{zhu2018hidden}, \cite{zhang2020udh} \\

Regeneration Attack & \cite{zhao2024invisible} & \ding{192}, \ding{194}& \cite{navas2008dwt}, \cite{zhang2019robust}, \cite{fernandez2022watermarking}, \cite{tancik2020stegastamp} \\

Transfer Evasion Attack  & \cite{hu2024transfer}&  \ding{192}, \ding{194} & \cite{zhu2018hidden}, \cite{tancik2020stegastamp}, \cite{fernandez2023stable}, \cite{jiang2024certifiably} \\
\rowcolor{gray!10}

Steganalysis & \cite{yang2024can} & \ding{192}, \ding{193} & \cite{al2007combined}, \cite{navas2008dwt}, \cite{zhu2018hidden}, \cite{zhang2019robust}, \cite{fernandez2022watermarking}, \cite{fernandez2023stable}, \cite{wen2023tree}, \cite{ci2024ringid}, \cite{yang2024gaussian}, \cite{xian2024raw}, \cite{ciwmadapter} \\

Imprint-Removal Attack & \cite{muller2025black} & \ding{192}, \ding{194}, \ding{196} & \cite{wen2023tree}, \cite{yang2024gaussian} \\

UnMarker  & \cite{kassis2025unmarker} & \ding{192} & \cite{tancik2020stegastamp}, \cite{lukas2023ptw}, \cite{zhu2018hidden}, \cite{fernandez2023stable}, \cite{wen2023tree}, \cite{yu2021artificial}, \cite{yuresponsible} \\

%TRW, Yu1, Yu2

CtrlRegen  & \cite{liu2024image} & \ding{192}, \ding{194} &  \cite{zhang2019robust}, \cite{fernandez2022watermarking}, \cite{fernandez2023stable}, \cite{tancik2020stegastamp}, \cite{wen2023tree} \\

CtrlRegen+  & \cite{liu2024image} & \ding{192}, \ding{194} & \cite{zhang2019robust}, \cite{fernandez2022watermarking}, \cite{fernandez2023stable} \\

  \midrule
  \rowcolor{red!10} 
\multicolumn{4}{c}{\textbf{Forgery attack}}\\
\midrule
Steganalysis & \cite{yang2024can} & \ding{192}, \ding{193} &\cite{wen2023tree}\\

Adversarial Forgery Attack & \cite{jain2025forging} & \ding{194}, \ding{192} & \cite{ci2024ringid}, \cite{arabi2024hidden},\cite{yang2024gaussian}, \cite{wen2023tree} \\

Imprint-Forgery Attack & \cite{muller2025black} & \ding{192}, \ding{194}, \ding{196} & \cite{wen2023tree}, \cite{yang2024gaussian} \\

Reprompting Attack & \cite{muller2025black} & \ding{192}, \ding{194}, \ding{196} & \cite{wen2023tree}, \cite{yang2024gaussian} \\

PnP  Universal Forgery & \cite{zhu2025optimization} & \ding{192}, \ding{193}, \ding{194}, \ding{196} & \cite{wen2023tree}, \cite{yang2024gaussian}\\

DiffForge & \cite{dong2025imperceptible} & \ding{192}, \ding{194} &\cite{zhu2018hidden}, \cite{zhang2019robust}, \cite{fernandez2023stable}\\

 Counterfeit Extractor Attack &\cite{luo2025watermark} & \ding{192}, \ding{193}& \cite{fernandez2023stable}, \cite{lei2024diffusetrace}, \cite{feng2024aqualora}\\

\bottomrule
\end{tabular}
}
\end{table*}

\subsection{Common Image Attacks}

Common image attacks refer to standard signal processing operations applied to watermarked images with the intent to degrade or remove embedded messages or watermarks, thereby reducing the detection score of watermark extraction. In most prior studies~\cite{fernandez2022watermarking,xiong2023flexible,kim2024wouaf,wang2024sleepermark,wen2023tree,ci2024ringid,zhang2024attack,yang2024gaussian,yang2025gaussian,gunnPRC,shukla2025waterflow,lu2024robust}, such attacks are primarily employed to evaluate the fundamental robustness of watermarking systems.

Common image attacks include \textit{geometric transformations} (such as rotation, scaling, cropping, translation, and resizing), \textit{noise perturbations} (such as Gaussian noise and salt-and-pepper noise), and \textit{lossy compression} (most notably JPEG), all of which can introduce irreversible distortions that compromise watermark detectability. Additionally, \textit{blurring} operations (such as Gaussian and median blur) can smooth out high-frequency components, potentially distorting or removing watermark signals. Even seemingly minor brightness and contrast adjustments may significantly impair watermark schemes that rely on subtle pixel-level intensity variations.

% These attacks can be broadly categorized into several types:

% \begin{itemize}
%     \item [(1)]\textbf{Geometric Attacks:} These involve spatial transformations such as rotation, scaling, cropping, translation, and resizing. Watermarking schemes embedded in the frequency domain are often specifically designed to maintain robustness against such geometric distortions.
%     \item [(2)]\textbf{Noise Perturbations:} This attack includes the injection of noise into the image, such as Gaussian noise or salt-and-pepper noise. These perturbations are often used to evaluate the noise resilience of watermarking methods.
%     \item [(3)]\textbf{Compression:} Lossy compression methods, particularly JPEG, are among the most common real-world transformations. Compression can introduce irreversible changes to pixel values, posing a major challenge to watermark detectability during storage or transmission.
%     \item [(4)] \textbf{Blurring:} Filters such as Gaussian blur and median blur can smooth out high-frequency components of the image, potentially distorting or removing watermark signals, especially those embedded in texture-rich regions.
%     \item [(5)] \textbf{Brightness and Contrast Adjustments:} Even basic modifications to brightness and contrast can serve as effective attacks, particularly against watermarks that rely on subtle pixel-level changes.

% \end{itemize}

\subsection{Removal Attacks}\label{sec:removal_att}
Image watermark removal attacks aim to manipulate AI-generated images such that $\mathsf{Extract}$ recovers incorrect information or $\mathsf{Verify}$ fails to identify the presence of a watermark. These attacks effectively disrupt both watermark detection and user attribution. Early approaches relied on simple, image-level transformations that only modestly reduced the detection score. However, more advanced adversaries have since developed specialized removal techniques capable of eliminating watermark traces while preserving high perceptual quality in the generated content. Formally, a watermark removal attack can be defined as:
\begin{equation}
\begin{aligned}
    &\min_{\delta} \|\delta\|_p \\ 
    \textit{s.t.}\;\mathsf{Verify}(\hat m, m)\rightarrow false&, \;\hat m= \mathsf{Extract}(I_w + \delta;K,\theta),\\
    %d_{\mathrm{vis}}(I, I_w) \le \delta 
%&\quad\text{and}\quad
%d_{\mathrm{sem}}\!\big(y, \phi(I_w)\big) \le \eta,
\end{aligned}
\end{equation}
where the adversary aims to fool $\mathsf{Verify}$ by manipulating $I_w$. $\delta$ denotes the perturbation adopted to $I_w$, and $\|\delta\|_p$ represents the $L_p$ norm measuring its magnitude. 

In this subsection, we discuss several representative watermark removal attacks that illustrate the main research directions in this area. 

\subsubsection{Regeneration Attacks}\label{sec:rengen_att} They represent a powerful class of watermark removal techniques. They aim to eliminate the watermark signal by regenerating the image through a non-watermarked generative model, while preserving its semantic fidelity. For instance, Zhao \textit{et al.}~\cite{zhao2024invisible} proposed a VAE attack, which leverages a VAE to map a watermarked image \( I_w \) into a latent representation and then reconstructs it as \( I' \), such that \( I' \approx I_w \). The success of the attack is based on that this reconstruction process is lossy due to the stochastic sampling and capacity limitations of the VAE, often resulting in subtle perceptual degradation, such as blurring, and this attack $A$ can be defined as:
\begin{equation}
I' = \mathcal{D}(\mathcal{E}(I_w)).
\end{equation}

Similarly, as illustrated in \Cref{eq:diff_att}, a diffusion attack was also proposed in~\cite{zhao2024invisible}, and this attack regenerates the image using a surrogate LDM different from the original $\mathsf{Gen}$, thereby altering the latent representation of $I_w$. This is achieved by adding Gaussian noise to the $\mathbf{z}_t$, which effectively degrades or removes the embedded watermark signal. In their experiments, the target generative model is SD-2.1, while SD-1.4 is used as the surrogate regeneration model.
\begin{equation}\label{eq:diff_att}
\begin{aligned}
    &I' = \mathcal{D}\big(\mathsf{DDIM}_{t\rightarrow 0}(\mathsf{DDIM}_{0\rightarrow t}(\mathcal{E}(I_w)) + \delta_{\mathbf{z}_t})\big), \\
    &\textit{where}\; t\in\{0,\dots,T\},\;\{\mathcal{D},\mathcal{E}\}\in\mathsf{Gen}^{\text{surrogate}}.
\end{aligned}\; 
\end{equation}

Different from Zhao \textit{et al.}'s approach~\cite{zhao2024invisible} in which regeneration begins from $\mathbf{z}_0$, Liu \textit{et al.} proposed CtrlRegen~\cite{liu2024image}, a watermark removal attack on the generative models that regenerates the image from a clean Gaussian noise $\mathbf{z}_T$ containing no embedded watermark information. To ensure that the regenerated image remains semantically and structurally consistent with the original watermarked input, CtrlRegen incorporates two control modules that guide the denoising process of the LDM. Furthermore, they proposed an enhanced variant, CtrlRegen$+$, which allows partial feature retention by injecting controllable noise into the latent representation of the watermarked image, offering a trade-off between watermark removal strength and visual consistency.

\subsubsection{Adversarial Attacks}\label{sec:adv_att} They are the most extensively studied watermark removal techniques. The core idea is to introduce imperceptible perturbations to a watermarked image while maintaining high perceptual quality, thereby deceiving the detector into classifying it as unwatermarked. This process can be formulated as a constrained optimization problem. The goal is to find the smallest possible perturbation $\delta$ that causes the $\mathsf{Verify}$ to identify the image as unwatermarked incorrectly. Such attacks are applicable at different levels of adversary prior knowledge and have demonstrated strong effectiveness in bypassing watermark detection systems. 

A representative example of adversarial watermark removal is WEvade, proposed by Jiang \textit{et al.}~\cite{jiang2023evading}. WEvade introduces an adversarial attack framework that enables watermarked images to evade detection by dual-tail watermark detectors while maintaining high perceptual quality. The framework is shown to be effective under both white-box and black-box settings. In the white-box scenario, where the adversary has full access to the $\mathsf{Extract}$ and $\mathsf{Verify}$, the attack seeks to reduce the decoded message's BA to approximately 0.5, rendering it statistically indistinguishable from that of a non-watermarked image. In the black-box setting, where only detection outcomes can be queried, the adversary constructs a surrogate model and employs an optimized query strategy to efficiently identify perturbations that evade detection.
Hu \textit{et al.}~\cite{hu2024transfer} extended this line of work to an even more challenging no-box setting, where the adversary has no access to the target watermarking model, neither its $\mathsf{Verify}$ and nor its API. To launch a transfer attack, the adversary trains an ensemble of surrogate watermarking models and crafts a single, optimized perturbation that successfully evades detection across all of them. The key assumption is that this adversarial perturbation will transfer to the unseen target model, enabling watermark evasion without direct access.

Yang \textit{et al.}~\cite{yang2024can} proposed a simple steganalysis-based watermark removal attack that extracts a content-agnostic, additive watermark pattern by averaging over a large collection of images, either paired (grey-box setting) or unpaired (black-box setting). By treating the watermark as an additive perturbation $\Delta$ consistent with the definition provided in \Cref{sec:term}, $\Delta$ can be subtracted from watermarked images to remove the watermark. This adversarial attack is effective against both fine-tuning-based and initial-noise-based watermarking methods. 

Imprint-removal attack~\cite{muller2025black} is also an adversarial method that removes initial-noise watermarks without requiring access to a generative model. Instead, the attacker uses a functionally unrelated surrogate model, such as a different version of SDM. The attack operates in the surrogate model's latent space by optimizing the target image’s $\mathbf{z}_0$ to disrupt the watermark signal embedded in $\mathbf{z}_T$. This method effectively removes watermarks from both TreeRing~\cite{wen2023tree} and Gaussian Shading~\cite{yang2024gaussian}.

Both regeneration attacks and adversarial attacks fall under the broader category of per-image attacks, in which each image is individually modified to remove or evade the watermark. In contrast, another class of attacks, known as model-targeted attacks, focuses on altering the generative model itself to consistently suppress or remove embedded watermark signals across all generated outputs. By fine-tuning model parameters, the adversary effectively overwrites or neutralizes the watermark-related weights within the LDM. Wu \textit{et al.}~\cite{wu2024robustness} demonstrated this approach by fine-tuning the image generator with a new watermark, effectively overwriting the original watermark signal. Similarly, Hu \textit{et al.}~\cite{hu2024stable} proposed a method in which the adversary retrains the VAE decoder using a set of unwatermarked images and their estimated latent representations. The fine-tuning objective is to ensure that the VAE decoder reconstructs images closely matching the original unwatermarked inputs when conditioned on these latent representations; this finding is also supported by~\cite{wang2024sleepermark}.

A recent state-of-the-art advancement is UnMarker~\cite{kassis2025unmarker}, a universal watermark removal framework that introduces distinct strategies for attacking post-hoc, fine-tuning-based, and initial-noise-based watermarking. For Stable Signature~\cite{fernandez2023stable}, where watermark signals are typically embedded in high-frequency components of generated images, UnMarker directly optimizes image pixels by maximizing the Fourier loss to disrupt the frequency spectrum. In TreeRing~\cite{wen2023tree}, where watermark signals are embedded in $\mathbf{z}_T$, they cannot be effectively removed through pixel-level operations. To address this challenge, UnMarker introduces adversarial filtering, which learns the weights of a set of trainable filters to systematically modify image structure and texture, thereby achieving effective TreeRing watermark removal.

%   1.  Distortion-Based Attacks
%   • Geometric attacks (cropping, rotation, resizing)
%   • Signal processing (blurring, compression, noise)

\subsection{Forgery Attacks}\label{sec:forgery_att}
Forgery attacks aim to embed a valid watermark into unmarked images to falsely claim authorship or ownership. Unlike removal attacks that erase identity, forgery seeks to deceive attribution by impersonating legitimate sources. Formally, a forgery attack is defined as:
\begin{equation}
\begin{aligned}
    &\min_{\delta} \|\delta\|_p \\ 
    \textit{s.t.}\;\mathsf{Verify}(\hat m, m)\rightarrow true&, \;\hat m= \mathsf{Extract}(I + \delta;K,\theta).\\
    %d_{\mathrm{vis}}(I, I_w) \le \delta 
%&\quad\text{and}\quad
%d_{\mathrm{sem}}\!\big(y, \phi(I_w)\big) \le \eta,
\end{aligned}
\end{equation}

M\"uller \textit{et al.}~\cite{muller2025black} proposed two forgery attacks against TreeRing \cite{wen2023tree} and Gaussian Shading \cite{yang2024gaussian}. The first attack, the Imprint-Forgery Attack, aims to transform the embedded watermark signal from a reference watermarked image $I_w$ onto any clean image $I$. Importantly, the attacker does not know the underlying key $K$ and message $m$; instead, it directly manipulates $\mathbf{z}_0$ so that the watermark signal associated with $I_w$ is inherited by the forged image $I'$. Specifically, the attacker estimates the watermarked initial noise $\hat{\mathbf{z}}^{w}_{T}$ of $I_w$ using a surrogate model, and then optimizes a perturbation $\delta_{\mathbf{z}_0}$ so that the inverted initial noise of $I'$ matches this target watermark signal:
\begin{equation}\label{eq:imprint}
\begin{aligned}
    &\hat{\mathbf{z}^w_T}  = \mathsf{DDIM}_{0\rightarrow T}(\mathcal{E}(I_w)),\quad \mathbf{z}_0  = \mathcal{E}(I),\quad \mathcal{L}(\delta_{\mathbf{z}_0})  = \|\mathsf{DDIM}_{0\rightarrow T}(\mathbf{z}_0 + \delta_{\mathbf{z}_0}) - \hat{\mathbf{z}^w_T}\|_2, \\ &I' = \mathcal{D}(\mathbf{z}_0 + \delta_{\mathbf{z}_0}), \quad \textit{where}\;\{\mathcal{D},\mathcal{E}\}\in\mathsf{Gen}^{\text{surrogate}}.
\end{aligned}
\end{equation}

As a result, $I'$ contains the same watermark signal as $I_w$ and thus yields the same extracted message under $\mathsf{Extract}$.

The second attack, the Reprompting Attack, does not manipulate a clean image but instead generates an entirely new image that inherits the watermark signal of $I_w$. The attacker extracts the initial noise $\hat{\mathbf{z}}^{w}_{T}$ from the watermarked image and directly reuses it as the starting noise for a surrogate model conditioned on a new prompt $y^{\text{new}}$. Since the watermark signal remains encoded in $\hat{\mathbf{z}}^{w}_{T}$, the newly synthesized image $I'$ carries the same embedded watermark and therefore leads to the same decoded message:
\begin{equation}\label{eq:reprompt}
\begin{aligned}
    \hat{\mathbf{z}}^{w}_{T} &= \mathsf{DDIM}_{0\rightarrow T}(\mathcal{E}(I_w)),\quad I'= \mathcal{D}\!\left(\mathsf{DDIM}_{T\rightarrow 0}(\hat{\mathbf{z}}^{w}_{T};\, y^{\text{new}})\right),
\end{aligned}
\end{equation}
where $\mathcal{D}$ and $\mathcal{E}$ are the VAE decoder and encoder of the surrogate model, respectively, and $y^{\text{new}}$ denotes the new prompt. This attack effectively performs a replay of the watermark signal while freely altering the semantic image content, revealing that the watermark is bound to the noise latent rather than the semantic prompt or the rendered pixels.

Dong \textit{et al.}~\cite{dong2025imperceptible} proposed DiffForge, a two-stage watermark forgery framework designed to embed a forged watermark signal into clean images. In the first stage, an unconditional LDM is trained on a collection of victim-generated, watermarked images to learn the distinct statistical and structural characteristics of the target watermark signal. In the second stage, DiffForge performs a shallow DDIM inversion on a clean target image, reversing it to a latent representation that preserves the original content while enabling watermark injection. The pre-trained diffusion model is then used to denoise this latent representation, thereby embedding the forged watermark naturally during the reconstruction process. 

Jain et al.~\cite{jain2025forging} proposed a more challenging single-sample forgery attack, requiring only one watermarked image. Exploiting the many-to-one mapping from latent space to noise, the adversary identifies vulnerable regions where benign images may be misclassified as watermarked. Rather than replicating the watermark, a small imperceptible perturbation is optimized to align the latent representation of a clean image with that of a watermarked one, inducing false positives. The attack only relies on a surrogate VAE encoder, without access to the LDM’s U-Net or original data, making it both practical and stealthy.

\subsection{Discussion}
The aforementioned attacks present distinct challenges that align with the two fundamental requirements of any watermarking system: robustness and security, as defined in \Cref{sec:fund_wm}.

Robustness reflects a watermark’s ability to resist signal processing and removal attempts. Basic robustness is evaluated through common manipulations like compression, geometric distortions, and noise~\cite{zhu2018hidden,xiong2023flexible,kim2024wouaf,wang2024sleepermark}, which typically reduce detection accuracy but seldom erase the watermark entirely. Advanced robustness faces deliberate removal attacks, such as UnMarker~\cite{kassis2025unmarker} and VAE regeneration~\cite{zhao2024invisible}, which are more effective but may compromise visual quality or efficiency. The WAVES benchmark~\cite{an2024waves} standardizes evaluation under such targeted conditions.

Security, on the other hand, concerns the system’s ability to resist misuse and forgery. The most critical threat in this domain is the forgery attack, which seeks to embed a valid watermark into an originally unmarked image to falsify ownership and mislead attribution mechanisms. Attacks such as DiffForge~\cite{dong2025imperceptible} and the one-shot forgery method~\cite{jain2025forging} exploit the many-to-one mappings inherent in LDMs to manipulate clean images into watermarked regions recognized by detectors. The clear distinction between robustness and security underscores the evolving threat landscape of AI-generated image watermarking: watermarking systems must be not only robust against degradation and removal but also secure against forgery and malicious misuse.
% In this subsection, we discuss the varied spectrum of attacks on AI-generated image watermarking, classifying them into three main categories. First, we analyze the attacker's knowledge and ability. Then, we examined common image attacks. Secondly, we investigated more sophisticated watermark removal techniques that seek to remove the watermark while maintaining image fidelity. These are further divided into per-image attacks, such as regeneration-based methods that use clean generative models and adversarial attacks that craft minimal perturbations to fool detectors, as well as model-targeted attacks that fine-tune the detector. Finally, we discussed forgery attacks, which represent a different threat focused on maliciously embedding a legal watermark into the unwatermarked image to mislead attribution systems. In conclusion, these attacks underscore the evolving security challenges that AI-generated watermarking systems face.

\section{Secure and Robust AI-generated Image Watermarking}\label{sec:robustwm}
In this section, we focus on strategies that aim to enhance the robustness and security of the watermarking system. Robustness has historically been a fundamental aspect of watermarking research, denoting the system's capacity to maintain watermark integrity across diverse attacks. Security essentially pertains to a watermark's capacity to withstand forging or misuse. We emphasize various exemplary strategies that have been suggested to tackle these difficulties.

\subsection{Data Augmentation}
The use of noise layers or distortion layers in watermarking is a common data augmentation strategy to enhance the robustness of watermarks. These methods typically employ end-to-end architectures, where the noise layer is specifically designed to be differentiable. This differentiability allows the watermark encoder and extractor to be jointly trained to adapt to various common image attacks, thereby making the watermark resilient to these specific challenges. Common attack types simulated by these layers include lossy compression (e.g., JPEG), various forms of noise (e.g., Gaussian noise), blurring, brightness, and contrast adjustments, and geometric transformations (e.g., rotation, cropping). Previous work, such as HiDDeN~\cite{zhu2018hidden}, initiates this methodology by employing a differentiable noise layer for imitating attacks and training to enhance image reconstruction, watermark reconstruction, and adversarial discrimination. Other methods involving noise layers have been proposed~\cite{fernandez2023stable,xiong2023flexible, yang2024gaussian,yang2025gaussian,lu2024robust}, confirming their effectiveness in improving robustness.

In addition, to defend against more complex watermark removal attacks, Gan \textit{et al.}~\cite{gan2025genptw} extended the design of the distortion layer in their latest work. Beyond incorporating common image attacks, their approach actively simulates AIGC editing during training to enhance generalization and improve the final watermarking model’s robustness. This includes \textit{``inpainting''} to simulate local content regeneration, \textit{``VAE reconstruction edition''} to mimic global semantic rewriting, and \textit{``watermark region removal''} to emulate malicious erasure. This highlights the growing need within the research community to improve watermark robustness against advanced attacks.

\subsection{Redundancy Coding}
Redundancy Coding (RC), corresponding to the $\mathsf{Code}$ process defined in \Cref{sec:comm_def}, encodes a message $m$ into a codeword $w$. It is a classical technique that introduces structured redundancy into the message, thereby enhancing the reliability of information transmission and recovery. Within AI-generated image watermarking, RC serves as an effective tool to enhance robustness. Despite its theoretical advantages, the use of RC in watermarking remains limited. Most watermarking schemes have inherently low capacity, typically only a few bits, so adding redundant bits for error correction further reduces the usable payload. Consequently, RC often becomes an impractical trade-off, and only a few recent methods have incorporated it into their designs.

From a robustness perspective, RC mitigates perturbations and data loss by encoding the message with error-correcting codes, enabling accurate recovery even when some bits are corrupted. Gaussian Shading~\cite{yang2024gaussian} implicitly adopts a form of redundancy by replicating the message multiple times to construct the codeword. During extraction, Gaussian Shading applies majority voting across all copies to recover the final bit values. This voting-based mechanism acts as a hard-decision decoding process, providing strong resilience against stochastic noise, lossy reconstruction, and adversarial perturbations. Although simple, this redundant embedding effectively increases bit-level reliability and serves as an intuitive realization of RC principles.

Besides, Pseudorandom error-correcting code (PRC) was proposed in~\cite{christ2024pseudorandom} and used in~\cite{gunnPRC} as a redundancy coding technique for AI-generated image watermarking. First, the PRC encoder generates a PRC codeword consisting of symbols ($+1$ or $-1$) based on a given key. During the embedding phase, the system constructs the initial noise $\mathbf{z}_T$ and then sets the sign of each component to match the corresponding symbol in the PRC codeword while retaining the original magnitudes. In the extraction phase, the system also performs exact inversion to recover an approximation of $\mathbf{z}_T$ from the image. 

Gaussian Shading$++$~\cite{yang2025gaussian} extends the aforementioned ideas by introducing a more systematic, communication-theoretic design. It splits the terminal latent variable $\mathbf{z}_T$ into two channels: the PRC Channel and the GS Channel. The GS Channel carries the actual watermark, while the PRC Channel encodes a seed via PRC to drive the random process in the GS Channel.

%Gaussian Shading$++$~\cite{yang2025gaussian} proposed a performance-lossless and deployment-oriented watermarking framework for SDM based on~\cite{yang2024gaussian} and~\cite{gunnPRC}. The idea is a double-channel latent design, where the latent space is split into a PRC Channel and a GS Channel. The PRC Channel encodes a pseudorandom seed via PRC to drive the random process in the GS Channel, thus retaining pseudorandomness even under a fixed key and eliminating key-management overhead. The GS Channel embeds the actual watermark bits through diffusion and stream-cipher-based encryption, followed by distribution-preserving sampling to guarantee that the watermarked latents follow the same standard Gaussian distribution as non-watermarked ones. This design ensures performance-lossless generation, while a later soft-decision decoding mechanism under the AWGN channel model enhances robustness against parameter variations. By further incorporating public-key signatures (ECDSA), Gaussian Shading++ extends to third-party verifiable settings, achieving a secure and practically deployable watermarking scheme for diffusion models.

\subsection{Cryptography}
Existing watermarking methods introduce slight yet detectable statistical biases into images. A powerful adversary can analyze a large number of watermarked images to uncover such statistical regularities, thereby localizing, removing, or even forging the watermark \cite{an2024waves,kassis2025unmarker,muller2025black,yang2024can}. For example, Stable Signature~\cite{fernandez2023stable} leaves pronounced watermark signals in the high-frequency pixel domain~\cite{kassis2025unmarker}. Likewise, the TreeRing family watermarking embeds a spectral watermark pattern in the Fourier domain of $\mathbf{z}_T$. Under a public-algorithm assumption, a verifier holding the secret key can reliably detect the watermark. If, however, the design relies on keeping the watermark pattern parameters or the embedding process itself secret (rather than a key), the scheme would conflict with Kerckhoffs’s principle and be vulnerable to trivial detection/forgery/removal once the pipeline is disclosed. All of the above violate the imperceptibility requirement of secure watermarking as defined in \Cref{sec:wm_prop}.

WIND~\cite{arabi2024hidden} and SEAL~\cite{arabi2025seal} enhance the security of initial-noise watermarking at both key and semantic levels. WIND uses a cryptographic $\mathsf{hash}$ with private salt to generate pseudo-random initial-noise keys and employs a two-stage design where Fourier-domain patterns act as group identifiers. Its security theorem ensures that recovering one key does not enable forging others, achieving unforgeable key-based security. SEAL extends this by binding watermarks to image semantics: it applies $\mathsf{SimHash}$ to image caption embeddings and combines the output with a patch index and secret salt to generate patch-wise noise. Verification requires consistency in both noise and semantics, which thwarts forgery and enables localization of tampered regions.

In Gaussian Shading~\cite{yang2024gaussian}, Yang \textit{et al.} address this problem for the first time, introducing a key cryptographic primitive, a stream cipher (specifically ChaCha20), to apply encryption to the redundantly encoded message, masking the repeated-copy operations; on this basis, they sample the initial noise via their distribution-preserving sampling algorithm. Finally, unlike TreeRing, even if the embedding process is exposed, the watermark signal pattern within $\mathbf{z}_T$ remains both invisible and unrecoverable to the adversary.

Imperceptibility has stronger guarantees in~\cite{gunnPRC}. PRC itself is a cryptographic primitive built upon a deeper assumption, the Learning Parity with Noise (LPN) problem, which is assumed to be hard even for subexponential-time algorithms~\cite{christ2024pseudorandom}. This watermarking design, grounded in this mature cryptographic assumption, provides theoretical guarantees for the watermark’s security. Consequently, without the key driven the generation of pseudorandom $c_K(w)$, it is computationally infeasible for an adversary to add a watermark to an unwatermarked image or to forge one. The authors also demonstrate, in Gaussian Shading \cite{yang2024gaussian}, when the stream-cipher key and the watermark are fixed, a ResNet18~\cite{he2016deep} classifier can successfully distinguish clean images from watermarked images violating imperceptibility.

Gaussian Shading$++$ enhances watermark security by combining \cite{gunnPRC} and \cite{yang2024gaussian}. Specifically, the PRC Channel employs pseudorandom error-correcting codes. Meanwhile, the GS Channel applies a stream cipher, where the seed is protected through a private-key–dependent hash, ensuring that the encrypted message exhibits randomness across executions. The concatenation of these two pseudorandom components forms a ciphertext that satisfies $\mathrm{IND}\${}\text{-}\mathrm{CPA}$ security. Besides, to defend against forgery attacks that exploit the reuse of initial noise, the method embeds both the user identifier (User ID) and its corresponding ECDSA digital signature, which is the first introduced in this field, into the GS Channel, enabling authentication during extraction through verification with the public key. This design ensures that, without access to the private signing key, an adversary cannot forge a valid watermark even when possessing knowledge of the embedding process.

\subsection{Discussion}
      Current methods for enhancing robustness and security in AI-generated image watermarking primarily rely on data augmentation, RC, and cryptography. These methods employ distinct design philosophies to achieve robustness and security. In terms of robustness, VINE~\cite{lu2024robust} and Stable Signature~\cite{fernandez2023stable} employ proactive training strategies, incorporating simulated attacks such as blurring and cropping to enhance the watermark's robustness. SleeperMark~\cite{wang2024sleepermark} addresses the challenge of model fine-tuning by designing a mechanism intended to disentangle the watermark from the model's semantic knowledge. In contrast, Gaussian Shading, Gaussian Shading$++$~\cite{yang2025gaussian}, and PRC~\cite{gunnPRC} root their robustness in the embedding mechanism itself. The former achieves this by deeply binding the watermark with the image's semantics, while the latter leverages the inherent error-correction properties of pseudorandom codes.

    Regarding security, the approaches range from procedural safeguards to cryptographic guarantees. The Stable Signature's security relies on keeping the watermark extractor private to defend against white-box attacks~\cite{fernandez2023stable}. Taking a more rigorous approach, Gaussian Shading$++$ \cite{yang2025gaussian} and PRC \cite{gunnPRC} utilize advanced cryptographic techniques to prevent forgery. Gaussian Shading$++$ integrates ECDSA public-key signatures to enable verifiable, third-party attribution even when model components are exposed. Ultimately, PRC provides the strongest security foundation, and its cryptographic undetectability fundamentally raises the barrier for an adversary to spoof or forge the watermark without possessing the key.

    Although aforementioned strategies have increasingly taken robustness and security into account, the results summarized in the \Cref{tab:existingAIwatermark} indicate that, aside from Gaussian Shading$++$ \cite{yang2025gaussian}, most existing methods still lack sufficient security guarantees. Moreover, due to the evolving nature of attack approaches, robustness remains inherently limited. While initial-noise-based methods typically exhibit stronger security by design, they are still vulnerable to Reprompting forgery attacks \cite{muller2025black}, largely due to the open-source nature of SDM and DDIM inversion pipelines. Additionally, PRC \cite{gunnPRC} has been shown to be highly fragile to basic image cropping and resizing \cite{francati2025coding}. These limitations underscore the fact that designing watermarking schemes for AI-generated images that simultaneously ensure robustness and security remains a fundamentally open problem and a critical direction for future research.

\section{Open Questions and Future Work} \label{sec:fu_work}

% \begin{tcolorbox}[colback=gray!5!white,
%                   colframe=gray!75!black,
%                   coltitle=white,
%                   title=Open Questions,
%                   fonttitle=\bfseries,
%                   arc=3mm,      % 控制圆角大小
%                   boxrule=0.8pt % 边框粗细
%                   ]
% \begin{enumerate}
%     \item RQ1: How can the detector avoid false positives for vanilla images?
%     \item RQ2: What trade-offs arise between detection accuracy and efficiency?
%     \item RQ3: How does the detector generalize across generative models?
% \end{enumerate}
% \end{tcolorbox}

AI-generated image watermarking is a rapidly evolving research field that still presents numerous open challenges and promising directions for future exploration. This section discusses four open questions that warrant further research and consideration.

%(1) developing efficient, high-fidelity watermarking robustness assessment methods to establish more realistic robustness evaluation benchmarks for the existing AI-generated image watermarking technologies; (2) designing structured and informative watermarks that combine metadata (such as model identifiers and timestamps) with cryptographic mechanisms like digital signatures to enhance their practicality and trustworthiness; (3) exploring watermarking schemes that support third-party public verification, allowing the public or regulatory bodies to independently verify content provenance without access to core secrets, thereby addressing the trust issues associated with centralized verification; and (4) developing multi-stakeholder watermarking technologies that enable multiple parties (such as model providers, users, and content editors) to embed their respective identity information throughout the image lifecycle, adapting to complex real-world application scenarios.

\subsection{Efficient, High-Fidelity Watermarking Robustness Assessment}
As discussed in \Cref{sec:attack}, existing state-of-the-art watermark removal attacks primarily rely on two technical paradigms: image regeneration using large generative models and iterative adversarial optimization to identify effective perturbations. While these approaches have revealed critical weaknesses in current watermarking systems, they remain limited by high computational cost and notable degradation of perceptual fidelity. Furthermore, the robustness of initial-noise-based watermarking schemes has yet to be comprehensively studied, leaving an incomplete understanding of their real-world resilience under practical conditions. A promising direction for future research is the development of efficient, high-fidelity robustness assessment frameworks that enable systematic evaluation of watermark resilience across diverse degradation scenarios. Instead of relying solely on computationally expensive stress testing, such frameworks should balance quantitative rigor with perceptual realism, for instance, by incorporating lightweight architectures, hybrid frequency–spatial perturbation modeling, or learning-based surrogates that emulate real-world distortions in a cost-effective manner. Equally important is the creation of unified benchmarks and standardized evaluation pipelines that jointly consider watermark detectability, visual quality, robustness, and computational efficiency. These advancements would facilitate fair comparisons, holistic evaluation, and ultimately drive the development of secure and reliable AI-generated image watermarking systems.

\subsection{Structured and Informative Watermarking} 
Initial-noise-based watermarking methods, such as PRC~\cite{gunnPRC}, have demonstrated the capability to embed large-capacity watermarks robustly, up to 500 bytes, and even 2500 bytes under benign conditions. However, existing approaches rarely define a structured or semantically meaningful watermark format that assigns explicit roles to individual bits (e.g., model identifier, generation timestamp, or provenance metadata), thereby limiting their interpretability and practical utility in real-world provenance tracing. Furthermore, most current designs overlook fundamental cryptographic mechanisms such as digital signatures and hash-based authentication, which are critical for ensuring authenticity, non-repudiation, and tamper resistance. This raises an open question: How can we design a structured watermarking scheme that encodes semantically rich and verifiable information while maintaining robustness and perceptual quality? Achieving this goal requires balancing encoding capacity, verifiability, and imperceptibility, as well as exploring how cryptographic integration affects both the fidelity and resilience of embedded watermarks. Establishing standardized formatting rules, verifiable encoding protocols, and evaluation metrics will be essential for advancing structured, secure, and interpretable watermarking in generative AI.

\subsection{Third-Party Public Verifiable Watermarking}
Most existing generative watermarking systems operate under the assumption that only the GenAI provider or the end user controls the watermark extractor. While this centralized architecture enables tight integration between the generative model and the watermarking mechanism, it inevitably introduces a trust asymmetry in multi-party environments. In such settings, external stakeholders, such as content viewers, independent auditors, or regulatory authorities, lack access to the extractor, preventing them from independently verifying the authenticity or provenance of the generated images. This concentration of verification authority limits transparency and undermines accountability, as the power of attestation remains confined to a single entity. An important open question, therefore, arises: How can watermark verification be decentralized while preserving strong security and privacy guarantees? Future research should explore cryptographic and zero-knowledge watermarking protocols that allow third parties to statistically verify ownership or authenticity without requiring direct access to the watermarking key or model internals. Such designs would enhance transparency, strengthen public trust, and reduce potential misuse while preserving the confidentiality of the underlying generative systems.

\subsection{Multi-Stakeholder Watermarking} 
Most existing watermarking methods for AI-generated images focus on embedding a single piece of information for a specific purpose. While effective in limited settings, these methods fall short of meeting the complex and multidimensional security requirements of today’s digital landscape. In practice, the image generation pipeline typically involves three key stakeholders: the model provider, the model user, and the image editor. Each stakeholder plays a distinct role and has legitimate reasons to embed a watermark that reflects their responsibility and contribution throughout the image’s lifecycle. This underscores the need for multi-stakeholder watermarking methodologies that can embed multiple, coexisting watermarks within a single image to represent all involved parties. A key open challenge lies in designing secure and robust multi-entity watermarking systems that maintain an optimal balance between watermark detectability and visual fidelity—an issue that remains crucial and warrants further in-depth exploration in future research.

\section{Conclusion} \label{sec:con}
In this paper, we presented a comprehensive and systematic review of watermarking techniques for AI-generated images, establishing a unified formalization of watermarking systems including their core components and fundamental properties, and offering a in-depth overview and critical analysis of current advancements. We categorized existing methods into two dominant paradigms (fine-tuning-based and initial-noise-based watermarking) and examined their underlying principles, capabilities, and limitations. Through this exploration, we underscored watermarking’s pivotal role in protecting IP, ensuring content authenticity, and building trust within the GenAI ecosystem. This survey also highlighted persistent challenges, including robustness and security threats such as removal and forgery attacks, and reviewed emerging defenses such as data augmentation, redundancy coding, and cryptographic verification mechanisms. By consolidating these diverse perspectives, this work not only synthesizes the state of the art but also charts a roadmap for future research toward secure, interpretable, and verifiable watermarking frameworks that balance innovation with accountability. Ultimately, advancing watermarking technologies is fundamental to preserving creators’ rights, mitigating the risks of AI misuse, and ensuring the sustainable, ethical, and transparent evolution of GenAI.

%%%%%
% \begin{equation}
% \text{PSNR} = 10 \cdot \log_{10} \left( \frac{MAX_I^2}{\text{MSE}} \right),
% \text{MSE} = \frac{1}{HW} \sum_{i=1}^{H} \sum_{j=1}^{W} (I_{ij} - I_{w}_{ij})^2
% \end{equation}

% \begin{equation}
% \text{SSIM}(x, y) =
% \frac{(2\mu_x \mu_y + C_1)(2\sigma_{xy} + C_2)}
% {(\mu_x^2 + \mu_y^2 + C_1)(\sigma_x^2 + \sigma_y^2 + C_2)},
% \end{equation}

% \begin{equation}
% \text{LPIPS}(x, y) =
% \sum_{l} \frac{1}{H_l W_l} \sum_{h, w}
% \| w_l \odot (\hat{f}^l_{hw}(x) - \hat{f}^l_{hw}(y)) \|_2^2,
% \end{equation}

% \begin{equation}
% \text{CLIP}(I, T) =
% \frac{\phi_I(I) \cdot \phi_T(T)}
% {\|\phi_I(I)\|_2 \, \|\phi_T(T)\|_2},
% \end{equation}

\bibliographystyle{ACM-Reference-Format}
\bibliography{mybib}

\appendix

\end{document}